\documentclass[a4paper,amsmath,amssymb,aps,prd,nofootinbib,superscriptaddress]{revtex4-1}
\usepackage[utf8]{inputenc}
\usepackage{graphicx}
\usepackage{booktabs}
\usepackage[english]{babel}
\usepackage{lineno}
\usepackage{localdefinitions}
\usepackage{multirow}
\usepackage{xcolor,url}
\usepackage{array}
\usepackage{subfigure}

\AtBeginDocument{
\heavyrulewidth=.08em
\lightrulewidth=.05em
\cmidrulewidth=.03em
\belowrulesep=.65ex
\belowbottomsep=0pt
\aboverulesep=.4ex
\abovetopsep=0pt
\cmidrulesep=\doublerulesep
\cmidrulekern=.5em
\defaultaddspace=.5em
}

\usepackage{epstopdf}
\usepackage{colordvi}
\usepackage{epsfig,psfrag,rotating,soul}
\def\beqa{\begin{eqnarray}}
\def\eeqa{\end{eqnarray}}
\newcommand{\be}{\ensuremath{\beta}}
\newcommand{\al}{\ensuremath{\alpha}}

\allowdisplaybreaks
\renewcommand{\arraystretch}{1.5}
\psfrag{lk}{$l_k$}
\psfrag{lm}{$l_m$}
\psfrag{blk}{$\bar {l}_k$}
\psfrag{blm}{$\bar {l}_m$}

\begin{document}

\title{Signature of 2HDM at Higgs Factories}

\author{Wenhai Xie}
\affiliation{Theory Division, Institute of High Energy Physics, Chinese Academy of Sciences, Beijing 100049, China}
\affiliation{School of Physics Sciences, University of Chinese Academy of Sciences, Beijing 100049, China}
\author{R.\ Benbrik}
\affiliation{Laboratoire de Physique fondamentale et appliqu\'ee Safi, Facult\'e Polydisciplinaire de Safi, Sidi Bouzid, B.P. 4162, Safi, Morocco}
\author{Abdeljalil Habjia}
\affiliation{Sultan Moulay Slimane University, Polydisciplinary Faculty, Research Team in Theoretical Physics and Materials (RTTPM), Beni Mellal, 23000, Morocco. }

\author{Souad Taj}
\affiliation{Sultan Moulay Slimane University, Polydisciplinary Faculty, Research Team in Theoretical Physics and Materials (RTTPM), Beni Mellal, 23000, Morocco. }
\author{Bin Gong}
\affiliation{Theory Division, Institute of High Energy Physics, Chinese Academy of Sciences, Beijing 100049, China}
\author{Qi-Shu Yan}
\affiliation{School of Physics Sciences, University of Chinese Academy of Sciences, Beijing 100049, China}
\affiliation{Center for Future High Energy Physics, Chinese Academy of Sciences, Beijing 100049, China}

\date{\today}
\begin{abstract}
The full one-loop corrections, both the weak and QED corrections, to the
process $e^+ e^- \to Z \phi $ ($\phi=h^0,H^0$) in the two Higgs doublet model (2HDM) at the Higgs factories are
presented. Up to the $O(\alpha_{em})$ level, the virtual
corrections are evaluated by using the FeynArts/FormCalc packages. The real emission
corrections are computed using the Feynman Diagram Calculation (FDC) package and the collinear divergences are regularized by the
structure functions of an electron. Using the FeynArts/FormCalc and the FDC packages, we study the
corrections in the Standard Model (SM) and the 2HDM,
respectively. Gauge dependence arising in the normalization of mixing angles is removed by using the pinch technique. After taking into account experimental constraints from the current LHC data,
we propose four interesting benchmark scenarios for future colliders. By using these benchmark scenarios, we evaluate the deviation of
$\Delta\sigma(e^+ e^- \to Z \phi)$ from their SM values. We also examine Higgs boson decays $\phi\to
b\bar{b}$ and $\phi\to \tau^+\tau^-$, which can have large electroweak (EW) contributions from triple Higgs couplings which are absent in the SM.
It is found that for these benchmark scenarios, both EW and real emission corrections are sizeable
and could be measured at a future $e^+ e^-$ colliders such as the ILC, CLIC, and CEPC.
\end{abstract}

\maketitle

\section{Introduction}


A Higgs-like boson ($h$) has been discovered in the first Run of the Large Hadron Collider (LHC) in 2012 \cite{Chatrchyan:2012xdj,Aad:2012tfa}. Based on the data of Run-1, both ATLAS and CMS collaborations have established a Higgs boson mass with $m_{h}= 125.09\pm 0.21 (stat.)\pm 0.11 (syst.)$ GeV~\cite{Aad:2015zhl}. ATLAS and CMS also performed several Higgs coupling measurements to a precision around $10\%-20\%$, such as Higgs couplings to di-bosons $VV$ with $ V = W^\pm, Z, \gamma$ through a global fit with input from the measurement of Higgs decaying into two weak bosons, weak vector boson associated processes, and VBF processes \cite{Khachatryan:2016vau}. Very recently, ATLAS and CMS have also measured Higgs couplings to the fermions of the third generation over 5$\sigma$, via the process $ pp \to V h (h \to b {\bar b})$ \cite{Aaboud:2018gay,Sirunyan:2018kst}, via the VBF process $pp \to jj h (h \to \tau^+ \tau^-)$ \cite{Aaboud:2018pen,Sirunyan:2017khh}, and $pp\to tt h$ with combined Higgs boson decay final states \cite{Sirunyan:2018hoz,Aaboud:2018urx}. With Run-2 data, high precision for these measured Higgs boson couplings has been achieved \cite{Sirunyan:2018koj,Aad:2019mbh}. These measurements demonstrated that the SM is consistent with the current Higgs data.

At the LHC, due to the large theoretical uncertainties like the Parton Distribution Function (PDF) and large experimental background,
the precision for the measured Higgs boson couplings can reach  $4\%-6\%$ or so at 300 fb$^{-1}$ \cite{CMS:2018qgz} . In contrast, due to its clean environment, an $e^+e^-$ collider can perform precision measurements on the production and decay of the observed Higgs boson. For example, at  240 GeV $e^+e^-$ collider, the Higgs-strahlung process $e^+ e^- \to Zh$ is the dominant Higgs production channel, where the Higgs boson can be reconstructed by using the recoiled $Z$ boson via its leptonic decay \cite{GarciaAbia:1999kv}. When a sufficiently large dataset is accumulated, say 1 ab$^{-1}$, the projected precision for the vertex $ZZh$ can reach to $0.1\%$ or so \cite{CEPCStudyGroup:2018ghi}. For a center-of-mass energy of 240-250 GeV and an integrated luminosity of 250 fb$^{-1}$ for each of the two detectors in each year, a total ${\cal O}(10^6)$ Higgs bosons can be produced in a 10 year running (a total integrated luminosity can be estimated as 5 ab$^{-1}$), which can lead to a precision of those currently measured Higgs couplings around a percent level or less \cite{Asner:2013psa,Peskin:2012we,Blondel:2013rn,Belusevic:2012sg,CEPCStudyGroup:2018ghi}. At the international linear collider (ILC) experiments \cite{Baer:2013cma}, the luminosity of Higgs-factory run can be around 250 fb$^{-1}$ at $\sqrt{s}=250$ GeV. The production rate of Higgs bosons from the Higgs-strahlung process can be enhanced when polarized beams \cite{MoortgatPick:2005cw} are used. However, at  $\sqrt{s}=500$ GeV and/or $\sqrt{s}=800$ GeV, the vector-boson fusion process $WW \to h$ can be dominant \cite{MoortgatPick:2005cw} and ILC will have the capability to reach a precision around 1\% for most of Higgs couplings \cite{Moortgat-Picka:2015yla}. Correspondingly, precise theoretical predictions to the physical observables related to Higgs bosons, like cross sections and branching ratios, are required.
%
%

In this paper, we propose some benchmark scenarios of 2HDM after taking into account theoretical constraints as well as experimental restrictions from recent LHC data. By using these benchmark scenarios, we will study the effects of electroweak radiative corrections to the production process $e^+ e^- \to Z \phi$ $\phi=h^0, H^0$ and report  full next-to-leading order (NLO) calculations on the cross-section  in 2HDM by including both virtual corrections up to one-loop and real emissions of a photon. Our results are consistent with \cite{LopezVal:2010vk} within the same set of 2HDM parameters. The complete SM one-loop correction of $\sigma(e^+ e^- \to Zh_{SM})$ was first time calculated in \cite{Fleischer:1982af} then followed by works \cite{Kniehl:1991hk} and \cite{Denner:1992bc}, later on had been calculated using the GRACE system\footnote{GRACE is a system of program packages for automatic calculation, see https://minami-home.kek.jp} in~\cite{Belanger:2002ik} and we reproduced the results for comparison. We find that the EW corrections of new physics in our benchmark scenarios can be of order $-10\% \sim -20\%$, while the contribution of real emission has a positive sign. Four benchmark points are studied in this paper and  found that both EW and real emission corrections are sizeable, and could be the physical target at future $e^+ e^-$ colliders such as the ILC, CLIC, and CEPC.

One-loop corrections of the Higgs boson couplings have been investigated in various models. QCD and electroweak corrections of $hf\bar{f}$ were calculated in the SM in Refs. \cite{Sakai:1980fa,Inami:1980qp,Drees:1989du,Fleischer:1980ub,Kniehl:1991ze}. Similar calculations had been done in various models beyond the SM like in the minimal supersymmetric SM (MSSM). Due to its sizeable SUSY-QCD effects, it was observed that $hf\bar{f}$ couplings can receive a large corrections \cite{Dabelstein:1995js,Haber:2000kq,Guasch:2001wv}. Moreover, $hf\bar{f}$ couplings had been intensively studied in the 2HDM \cite{Kanemura:2014dja,Arhrib:2004ak,Arhrib:2016snv}. In this work, we also examine the deviations of $hf\bar{f}$ ($f=b,\tau$) in both type-I and type-II scenarios and find these deviations are within the reach of future Higgs factories and are helpful to distinguish new physics models.

%

The plan of the paper is as follows. In section 2, we introduce the benchmark points of the 2HDM model.
In section 3, we outline the framework of our calculation and specifies
the renormalization scheme we will use. In section 4, we present the numerical results.
We conclude this work in section 5.

\section{The 2HDM model and benchmark points}
2HDM was first introduced by T.D. Lee \cite{Lee:1973iz} in order to have spontaneous T parity (or equally CP parity) breaking, and a more comprehensive review can be found in Refs. \cite{Branco:2011iw} and \cite{Davidson:2005cw}. Before proposing benchmark scenarios, we briefly review the main features of 2HDM model related to this work\footnote{
In this work, we follow the notations of the reference \cite{Arhrib:2016snv} where more details can be found.}.
The most general renormalizable potential which is invariant under $SU(2)\times U(1)$ can be written as
\begin{eqnarray}
V(\Phi_1, \Phi_2) &=&\nonumber m^2_1 \Phi^\dagger_1 \Phi_1 + m^2_2 \Phi^\dagger_2 \Phi_2 - \left(m^2_{12} \Phi^\dagger_1 \Phi_2 + \mathrm{h.c.}\right) + \frac{1}{2}\lambda_1 (\Phi^\dagger_1 \Phi_1)^2 + \frac{1}{2}\lambda_2 (\Phi^\dagger_2 \Phi_2)^2 \\\nonumber &+& \lambda_3 (\Phi^\dagger_1 \Phi_1)(\Phi^\dagger_2 \Phi_2) + \lambda_4 (\Phi^\dagger_1 \Phi_2)(\Phi^\dagger_2 \Phi_1) \\ &+&
\left(\frac{1}{2}\lambda_5 (\Phi^\dagger_1 \Phi_2)^2 +\left(\lambda_6 (\Phi^\dagger_1 \Phi_1) + \lambda_7 (\Phi^\dagger_2 \Phi_2)\right)(\Phi^\dagger_1 \Phi_2) + \mathrm{h.c.}\right),
\label{eqa1}
\end{eqnarray}
where $\Phi_i$, $i=1,2$ are complex $SU(2)$ doublets with 4 degrees of freedom each and all $m^2_1$, $m^2_2$ and $\lambda_{1-4}$ are real which follows from the hermiticity of the potential. However, the parameters $\lambda_{5-7}$ and $m^2_{12}$ can be complex in general. Explicit CP-violation may arise in the Higgs sector by considering the imaginary parts of the above complex parameters. In this work we focus on our study in the CP-conserving case so that we assume all parameters to be real. Furthermore, for the sake of simplicity, we only include the soft breaking term proportional to $m_{12}^2$ and omit the terms proportional to $\lambda_6$ and $\lambda_7$ as these lead to hard $Z_2$ violation.

The $Z_2$ symmetry is defined as ($\Phi_1$, $\Phi_2) \to  (\Phi_1, -\Phi_2)$, which was introduced in order to suppress the tree-level flavor changing neutral current (FCNC) processes \cite{Glashow:1976nt,Paschos:1976ay}. 
The exact $Z_2$ symmetry will lead to the absence of  $\Phi_2$ couplings into fermions.
which makes $\Phi_2$ possible as a dark matter candidate in the so-called inert doublet model \cite{Ma:2006km,Barbieri:2006dq,Cao:2007rm}. However in this work, we consider the case $\lambda_6=\lambda_7=0$ but $m_{12}^2\neq 0$, which can allow FCNC processes at loop level \cite{Glashow:1976nt,Gunion:2002zf}. 

In CP-conserving 2HDM case, four realizations have been considered in the 2HDM to avoid FCNC at the tree-level, known as type I, type II, type III (called Lepton specific ) and type IV (also called Flipped) \cite{Barger:1989fj, Grossman:1994jb}. In the type I model, the $Z_2$ symmetry is an exact symmetry and the $\Phi_1$ doublet gives mass to all fermions. In the type II model, the $\Phi_1$ gives mass to leptons and down quarks while the  $\Phi_2$ couples to up-type quarks. The MSSM has a type II Higgs sector. In the type III model, all quarks couple to the $\Phi_2$ doublet while all charged leptons couple to the $\Phi_1$ doublet. Finally, in the type IV model,  down-type quarks couple to the $\Phi_1$ doublet while the rest  of fermions couple to the $\Phi_2$ doublet. With such four arrangements in flavor space of 2HDM, the tree-level flavour changing neutral current can be suppressed safely. 

From the initial eight degrees of freedom, if the $SU(2)$ symmetry is broken, we end up with two CP-even Higgs states usually denoted by $h^0$ and $H^0$, one CP-odd $A^0$, two charged Higgs boson, $H^\pm$ and three Goldstone bosons. After the electroweak symmetry breaking, the potential in Eq.~(\ref{eqa1}) can be expressed with 7 independent parameters namely $m_{h^0}$, $m_{H^0}$,$m_{A^0}$, $m_{H^\pm}$, $\tan\beta=v_2 / v_1$, $\sin(\beta -\alpha)$ and $\lambda_5$ (or equivalently $m^2_{12}$ ). The angle $\beta$ is the rotation angle from the weak gauge eigenstates to the mass eigenstates in the CP-odd and charged Higgs sector. The angle $\alpha$ is the corresponding rotation angle for the CP-even sector. From this potential, the triple Higgs couplings needed for the present work can be derived, which are functions of the physical parameters and are given in Eqs. (\ref{lll}-\ref{hpH}) of the Appendix \ref{trihiggsc}.\\

The parameter space of the 2HDM is reduced by the following theoretical and experimental constraints:
\begin{enumerate}
\item Vacuum stability conditions that ensure the potential is bounded from below, where we use the conditions derived in \cite{Deshpande:1977rw}.
\item Perturbative tree-level unitarity for scattering amplitudes of Higgs bosons and longitudinally components of gauge bosons.
\item The perturbativity of all quartic coefficients of the scalar potential , i.e $|\lambda_i| \leq 8 \pi$ ($i=1,...,5$),
\item EW Precision Observables (EWPOs). Due to the contributions of extra Higgs bosons, the universal parameters $S$, $T$ and $U$ provide additional constraints on the mass splitting between these Higgs bosons. To implement the constraints from EWPOs, we consider the following values \cite{Agashe:2014kda, Haller:2018nnx} for $S$, $T$ and $U$: $\Delta S = 0.05\pm 0.11$, $\Delta T = 0.09\pm 0.13$ and $\Delta U = 0.01\pm 0.11$.
\item Indirect experimental constraints from $B$~physics observables, which have been taken into account by using SuperIso \cite{Mahmoudi:2008tp}  public code. Several important experimental values are tabulated in Table \ref{Tab:ExpResult}, which will be used to constrain parameters such as $\tan\beta$ and the charged Higgs boson mass of the 2HDM significantly. 
\end{enumerate}

\begin{table}[th!]
	\begin{center}
		\begin{tabular}{|c|c|c|c|}
			\hline
			\hline Observable 		& Experimental result & SM contribution  & Combined error at 1$\sigma$\\
\hline $\mathcal B (K\to\mu\nu) / \mathcal B(\pi\to\mu\nu)$	& $0.6357 \pm 0.0011$~\cite{Agashe:2014kda} & $ 0.6231\pm 0.0071 $& 0.0071 \\
			\hline $\overline{\mathcal B}(b\to s\gamma)_{E_\gamma>1.6\,\text{GeV}}$ & $(3.32 \pm 0.16)\times 10^{-4}$~\cite{Amhis:2016xyh}& $(3.36\pm 0.24) \times 10^{-4}$ & $0.29\times 10^{-4}$\\
			\hline $\mathcal B (B\to\tau\nu)$									& $(1.14 \pm 0.22)\times 10^{-4}$~\cite{Amhis:2014hma}& $(0.78\pm 0.07 )\times 10^{-4}$ & $0.23\times 10^{-4}$\\
			\hline $\mathcal B (D\to\mu\nu)$ & $(3.74 \pm 0.17)\times 10^{-4}$~\cite{Agashe:2014kda,Amhis:2014hma}& $(3.94\pm 0.13)\times 10^{-4}$  & $0.21 \times 10^{-4}$\\
			\hline $\mathcal B(D_s\to\tau\nu)$ & $(5.55 \pm 0.24)\times 10^{-2}$~\cite{Agashe:2014kda,Amhis:2014hma}& $(5.17\pm 0.11) \times 10^{-2}$ & $0.26\times 10^{-2}$\\
			\hline $\mathcal B (D_s\to\mu\nu)$	& $(5.57 \pm 0.24)\times 10^{-3}$~\cite{Agashe:2014kda,Amhis:2014hma}& $(5.28\pm 0.11) \times 10^{-3}$ & $0.26\times 10^{-3}$\\
			\hline $\overline{\mathcal B} (B^0_s \to \mu^+ \mu^-)$ 				& $(2.8\pm0.7)\times10^{-9}$~\cite{Archilli:2014cla} &$(3.66\pm0.28) \times 10^{-9}$ & $0.75\times 10^{-9}$ \\
			\hline $\overline{\mathcal B} (B^0_d \to \mu^+ \mu^-)$& $(3.9\pm 1.5)\times 10^{-10}$~\cite{Archilli:2014cla}& $(1.08\pm 0.13) \times 10^{-10}$ & $1.50\times 10^{-10}$ \\
			\hline $\Delta M_s$ & $(17.757\pm 0.021)\text{ ps}^{-1}$~\cite{Amhis:2016xyh}&$(18.257\pm 1.505) \text{ ps}^{-1}$ & $1.5\text{ ps}^{-1}$\\
			\hline $\Delta M_d$ & $(0.510\pm 0.002) \text{ ps}^{-1}$~\cite{Amhis:2016xyh}	& $(0.548\pm0.075) \text{ ps}^{-1}$ & $0.075 \text{ ps}^{-1}$\\
			\hline $\Delta_0 (B\to K^*\gamma)$ 									& $(5.2\pm 2.6)\times 10^{-2}$~\cite{Agashe:2014kda}	& $(5.1\pm 1.5) \times 10^{-2}$ & $3.0\times 10^{-2}$\\
			\hline $\delta a_\mu$ & $(261\pm 80)\times 10^{-11}$~\cite{Hagiwara:2011af}	& $-$ & $80\times 10^{-11}$\\
			\hline
			\hline
		\end{tabular}\end{center}
		\caption{{
			Experimental results of the observables combined by the Particle Data Group (PDG) and/or Heavy Flavor Averaging Group
 (HFAG) Collaborations in Refs.~\cite{Agashe:2014kda}--\cite{Amhis:2014hma}.
			As for $\overline{\mathcal B} (B^0_q \to \mu^+ \mu^-)$, the combined results from the LHCb and CMS collaborations are shown as
given in Ref.~\cite{Archilli:2014cla}. Ref.~\cite{Hagiwara:2011af} is used for constraints from $(g-2)_\mu$ data.}
		}\label{Tab:ExpResult}
	\end{table}


The 2HDMC public code \cite{Eriksson:2009ws} allows us to check all the listed constraints above. Furthermore, the proposed benchmark scenarios in this paper satisfy two more constraints: 1) the limits obtained from various searches for additional Higgs bosons at the LHC and other collider data, and 2) by the requirement that there exists a neutral scalar which should match the measured properties of the Higgs-like boson. We evaluate the first constraint with the public code {\tt HiggsBounds-5.3.2}~\cite{Bechtle:2008jh,Bechtle:2011sb,Bechtle:2013gu,Bechtle:2013wla,Bechtle:2015pma}, and the second constraint with the code {\tt HiggsSignals-2.2.3}~\cite{Bechtle:2013xfa,Stal:2013hwa,Bechtle:2014ewa}.

Here we stress that after the Higgs-like particle discovery, several theoretical studies have performed global-fit analysis for the 2HDM to pinpoint the allowed regions of parameter space both for a SM-like
Higgs $h^0$ \cite{Ferreira:2011aa,Bernon:2015qea,Dumont:2014wha,Cheung:2013rva,Eberhardt:2013uba,Coleppa:2013dya,Arhrib:2015maa,Chiang:2013ixa} as well as for a SM-like Higgs boson $H^0$ \cite{Bernon:2015wef,Ferreira:2012my}. Before presenting our results, we would like to mention that we have performed a cross check
of the results ~\cite{LopezVal:2010vk} for the subclass of Yukawa
corrections considered there and found a perfect agreement.

\begin{figure}[htbp]
\setcounter{subfigure}{0}
 \centering
 \includegraphics[width=0.45\textwidth]{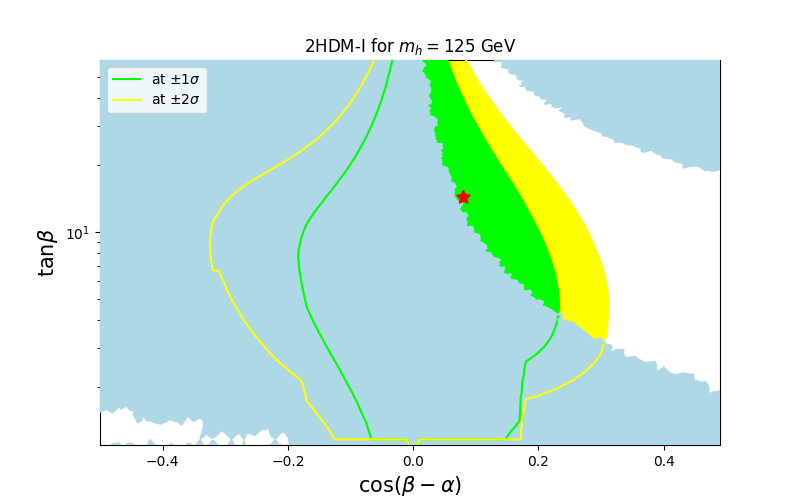}
 \includegraphics[width=0.45\textwidth]{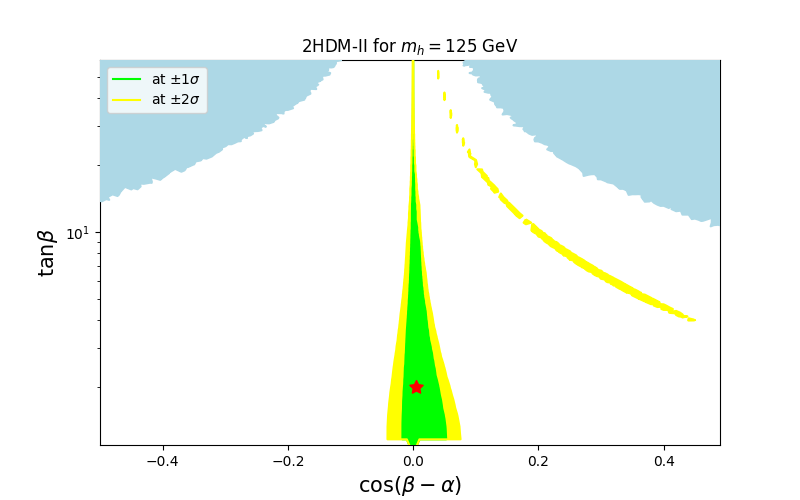}
\includegraphics[width=0.45\textwidth]{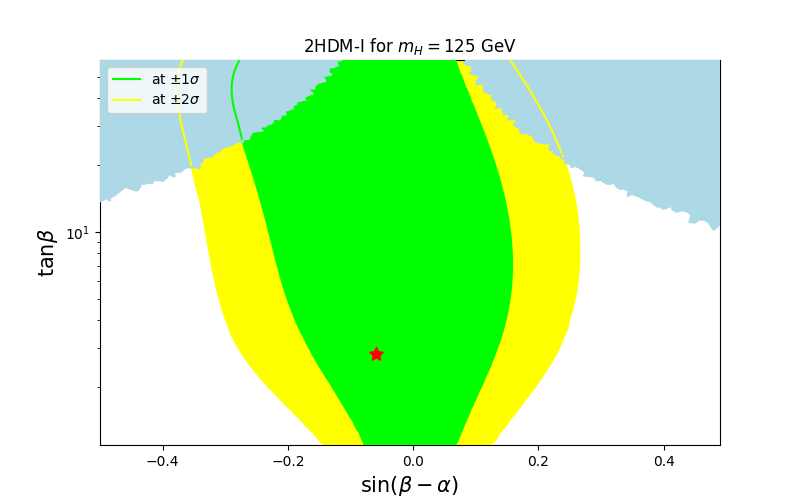}
\includegraphics[width=0.45\textwidth]{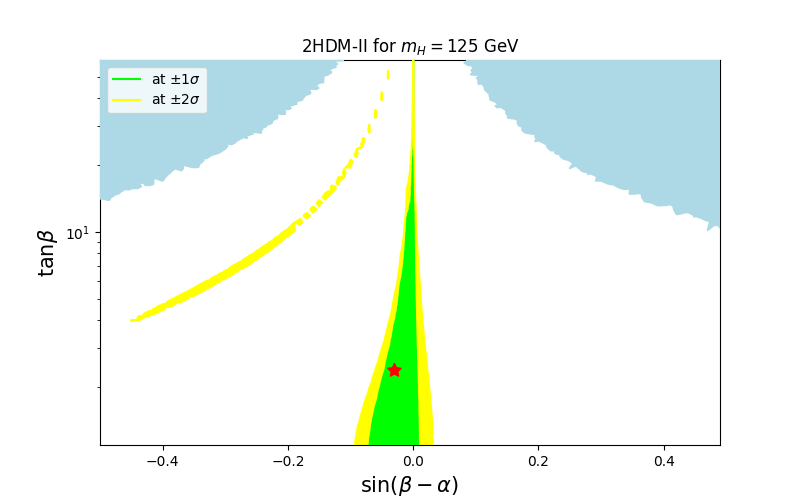}

 \caption{ Allowed ranges at 95\% CL (yellow color) and 68\% CL (green color) from direct search data at the LHC Run-II are shown. Excluded regions by theoretical constraints are given by cyan color. A red star  corresponds to the best fit point for each scenario. } \label{const}
\end{figure}

To explore the effect of the new physics in the 2HDM in our following study, we propose four benchmark points (BPs) as shown in Table \ref{BPdata} based on the best-fit from the latest results of Higgs data using HiggsBounds and HiggsSignals public codes.

\begin{table}[h!]
  \centering
  {\renewcommand{\arraystretch}{1.7} 
    {\setlength{\tabcolsep}{.21cm}
      \begin{tabular}{| c | c | c | c | c | c | c | c | c |}
        \hline
       BPs & $\sin(\beta-\alpha)$ & $ \tan \beta$ & $m_h$ (GeV) & $ m_H$ (GeV) & $m_A$ (GeV) &  $m_{H^\pm}$ (GeV) & $\lambda_5$ \\
        \hline
        \hline
        BP1-h & 0.99679 & 14.300 & 125.00 & 212.00 & 98.20 & 178.27 & 0.5819\\ \hline
        \hline
        BP2-h & 0.99999 & 2.012 & 125.00 & 594.00 & 512.00 & 592.00 & 0.0000\\ \hline
        \hline
         BP1-H & -0.06000 & 2.830 & 95.00 & 125.00 & 169.00 & 170.00 & -0.3220 \\ \hline
        \hline
         BP2-H & -0.03000 & 2.160 &95.00 & 125.00 & 600.00 & 600.00 & -5.7800 \\
        \hline
        \hline
      \end{tabular}}}
      \caption{Selected benchmark points using Higgs data at 13 TeV with $m_h$ = 125 GeV are presented, for BP1-h and BP2-h and with $m_H$ = 125 GeV, for BP1-H and BP2-H. Obviously, in this notation, BP1 (BP2) type is related to type-I (type-II). Here we adopt four physical masses, two angles, and $\lambda_5$ to describe each point in the parameter space. In each BP, we examine the ratio of total decay width over the mass of a Higgs boson, i.e. $\Gamma_\phi /m_\phi $, and find it is smaller than $5\%$ with $\phi = H, A$ and $H^\pm.$ }
\label{BPdata}
    \end{table}

To demonstrate the impact of experimental data as well as theoretical constraints upon the parameter spaces, we perform a scan on $\sin(\beta-\alpha)/\cos(\beta-\alpha)$ and $\tan\beta$ plane. The results are shown in Figure (\ref{const}) where we project the LHC constraints discussed above onto allowed regions at 95\% Confidence Level (CL) (yellow color) as well as at 68\% CL (green color) in the ($\sin(\beta-\alpha), \tan\beta$) plane for $h$ being SM-like for type-I (up left panel) and type-II (up right panel) 2HDM. In the lower plots, the heavier Higgs boson $H$ is the SM-like Higgs boson with the same coding color. The selected four benchmark points are given in Table \ref{BPdata}, each of which corresponds to the red star in each plot of Figure (\ref{const}). \\

About the mass of the charged Higgs boson, here we would like to point out that a light charged Higgs boson less than 200 GeV can be consistent with the current LHC Higgs data. As it is well known that in the framework of 2HDM-II and IV, for example, the measurement of the $b\to s\gamma$ branching ratio requires the mass of charged Higgs boson to be heavier than 580~GeV~\cite{Misiak:2017bgg,Misiak:2015xwa} for any value of $\tan\beta \geq 1$.
Such a limit is much lower for the other 2HDM types~\cite{Enomoto:2015wbn}. In 2HDM-I and III, as long as $\tan\beta\geq 2$, there are allowed regions in the parameter space with a charged Higgs bosons as light as 100 GeV \cite{Enomoto:2015wbn,Arhrib:2016wpw}, which are still consistent with all $B$ physics constraints as well as with LEP and LHC limits, as shown in the literature \cite{Aad:2014kga,Khachatryan:2015qxa,Khachatryan:2015uua,Aad:2013hla,Abbiendi:2013hk,Akeroyd:2016ymd}.

\begin{figure}[h]
 \centering
 \includegraphics[width=0.204\textwidth]{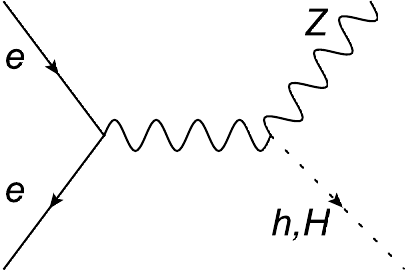}
 \caption{The LO Feynman diagram for the process $e^{+}e^{-}\to Zh$ and $e^{+}e^{-}\to ZH$ in the 2HDM is shown. } \label{lo}
\end{figure}

Due to the tiny Yukawa coupling of Higgs/Goldstone to electron, the overwhelming and leading order (LO) contribution for the process $e^{+}e^{-}\to Zh$ and $e^{+}e^{-}\to ZH$ in the 2HDM comes from the Feynman diagram given in Figure (\ref{lo}). 
Then the total tree-level cross-section in 2HDM $\sigma^0$ can be expressed as
\begin{equation}
\sigma^0\left(e^{+} e^{-} \rightarrow Z \phi\right) = \sin^2(\beta-\alpha)[\cos^2(\beta-\alpha)]\sigma^0_{SM}\left(e^{+} e^{-} \rightarrow Z h\right) \quad\quad {\rm for} \quad\quad \phi = h[H]
\end{equation}
where the tree-level cross section of the SM $\sigma^0_{SM}$ is defined as
\begin{equation}
\sigma^0_{SM}\left(e^{+} e^{-} \rightarrow Z h\right)=\frac{\alpha_{em}^2 \pi}{192 s s_W^{4}  c_W^{4} }\left(v_{e}^{2}+a_{e}^{2}\right) \lambda^{\frac{1}{2}} \frac{\lambda+12 m_{Z}^{2} / s}{\left(1-m_{Z}^{2} / s\right)^{2}}
\end{equation}
where $s$ is the collision energy, $a_{e}=-1, v_{e}=-1+4 \sin ^{2} \theta_{W}, \alpha_{em}={e^2}/{4 \pi},\lambda=\left(1-{m_{h}^{2}}/{s}-{m_{Z}^{2}}/{s}\right)^2- {4 m_{h}^{2} m_{Z}^{2}}/{s^{2}}$. Here we have used $s_W$ and $c_W$ for $\sin(\theta_W)$ and $\cos(\theta_W)$, and $\theta_W$ is the Weinberg angle.
With the relation $\alpha_{em}={\sqrt{2}m_W^2 s_W^2 G_F}/{\pi}$, it can be seen that our LO result in the SM is consistent with those in Refs.~\cite{Kilian:1995tr,Carena:1996bj,Djouadi:2005gj} .
The formula of tree-level cross sections for Higgs-strahlung process at $e^+ e^-$ colliders in the 2HDM can be also found in \cite{Djouadi:1992pu}, where the MSSM was considered which has a built-in type-II 2HDM in the Higgs sector.


\section{One-loop renormalization and calculation\label{sec:one-loop}}
Calculations of higher-order corrections in perturbation theory in general
lead to ultra-violet (UV) divergences.
The standard procedure to eliminate these UV divergences consists in the
renormalization of bare Lagrangian by redefining the couplings and
fields. In the SM, the on-shell renormalization scheme is
well elaborated~\cite{Bohm:1986rj,Hollik:1988ii,Denner:1991kt}.

As shown in Ref.~\cite{Denner:1991kt}, the renormalization constant for charge $\delta Z_e$ is obtained as
\begin{equation}
\delta Z_e=-\dfrac{1}{2}\delta Z_{AA}-\dfrac{s_W}{c_W}\dfrac{1}{2}\delta Z_{ZA}
=\dfrac{1}{2}\Pi(0)-\dfrac{s_W}{c_W}\dfrac{\sum^{AZ}_T(0)}{m_Z^2}
\end{equation}
with
\begin{equation}
\Pi(s)\equiv\dfrac{\sum^{AA}_T(s)}{s},
\end{equation}
and
\begin{equation}
\Pi(0)=\lim_{s\rightarrow 0}\dfrac{\sum^{AA}_T(s)}{s}=\dfrac{\partial\sum^{AA}_T(s)}{\partial s}\biggr|_{s=0}.
\end{equation}
where the definitions for $\delta Z_e$, $\delta Z_{AA}$ and $\delta Z_{ZA}$ can be found in Eqs.~(\ref{eqn:re_e}) and (\ref{eqn:re_za}). 
This corresponds to the running coupling constant obtained in the Thomson limit, taken as $\alpha_{em}(m_e)$ or $\alpha_{em}(0)$.  
The vacuum polarization $\Pi(0)$, the first term in $\delta Z_e$, is sensitive to the hadronic contribution. Usually a non-perturbative parameter $\Delta\alpha^{(5)}_{\mathrm{hadron}}(m_Z)$ (``5'' here is the number of quark flavors, which means top quark is not included) is used to absorb the hadronic contribution, i.e. $\Pi(0)$ is modified as (the contributions from leptons are also separated for further discussion):
\begin{equation}
\Pi(0)=\Pi_{\mathrm{hadron}}^{(5)}(0)+\Pi_{\mathrm{lepton}}(0) +\Pi_{\mathrm{remaining}}(0) \rightarrow \mathrm{Re}\Pi^{(5)}_{\mathrm{hadron}}(m_Z^2)+\Delta\alpha^{(5)}_{\mathrm{hadron}}(m_Z)+\Pi_{\mathrm{lepton}}(0) +\Pi_{\mathrm{remaining}}(0), 
\label{eqn:pi}
\end{equation}
with $\Delta\alpha^{(5)}_{\mathrm{hadron}}(m_Z)=0.02764$ according to PDG data~\cite{Tanabashi:2018oca}.
Then $\delta Z_e$ is rewritten as
\begin{equation}
\label{eqn:a0}
\delta Z_e(0)=\dfrac{1}{2}\mathrm{Re}\Pi^{(5)}_{\mathrm{hadron}}(m_Z^2)+\dfrac{1}{2}\Delta\alpha^{(5)}_{\mathrm{hadron}}(m_Z)+\dfrac{1}{2}\Pi_{\mathrm{lepton}}(0)+\dfrac{1}{2}\Pi_{\mathrm{remaining}}(0)-\dfrac{s_W}{c_W}\dfrac{\sum^{AZ}_T(0)}{m_Z^2} .
\end{equation}
In the following text and Table \ref{nlo_SM}, we label the scheme defined in Eq.~(\ref{eqn:a0}) as $\alpha_{em}(0)$ scheme. Meanwhile, whenever $\Pi(0)$ is mentioned, it refers to the one defined in Eq.~(\ref{eqn:pi}).

Another two schemes labelled as $\alpha_{em}(m_Z)$ and $\alpha_{em}(\sqrt{s})$ are defined below where the large logarithmic contributions of leptons are also absorbed into the redefinition of running coupling constant~\cite{Denner:1991kt,Sun:2016bel}
\begin{equation}
\delta Z_e(\mu)\equiv\delta Z_e(0)-\dfrac{1}{2}\Delta\alpha(\mu)=\dfrac{1}{2}\mathrm{Re}\Pi_{\mathrm{hadron}}^{(5)}(\mu^2) + \dfrac{1}{2}\mathrm{Re}\Pi_{\mathrm{lepton}}(\mu^2) 
+\dfrac{1}{2}\Pi_{\mathrm{remaining}}(0)-\dfrac{s_W}{c_W}\dfrac{\sum^{AZ}_T(0)}{m_Z^2}, 
\label{eqn:scale}
\end{equation}
with
\begin{eqnarray}
\Delta\alpha(\mu)&\equiv&\Pi_{f\neq \mathrm{top}}(0)-\mathrm{Re}\Pi_{f\neq \mathrm{top}}(\mu^2)
\nonumber \\
&=&[\mathrm{Re}\Pi^{(5)}_{\mathrm{hadron}}(m_Z^2)+\Delta\alpha^{(5)}_{\mathrm{hadron}}(m_Z)-\mathrm{Re}\Pi_{\mathrm{hadron}}^{(5)}(\mu^2)] 
+[\Pi_{\mathrm{lepton}}(0) -\mathrm{Re}\Pi_{\mathrm{lepton}}(\mu^2)]
\end{eqnarray}
And the running coupling constant is defined as 
\begin{equation}
\alpha_{em}(\mu)\equiv\dfrac{\alpha_{em}(0)}{1-\Delta\alpha(\mu)}.
\end{equation}

In the $\alpha_{em}(m_Z)$ scheme, we take $\mu = m_Z$, while in the $\alpha_{em}(\sqrt{s})$ scheme, we take $\mu = \sqrt{s}$. Results in these two schemes will be independent of $\log(m_e)$. In the following, we use the $\alpha_{em}(m_Z)$ scheme as our default choice. 
In the procedure of renormalization using dimensional regularization, a scale $\mu_r$ is introduced. Usually this will generate a logarithmic term $\log(\mu_r)$. 
As we use a complete on-shell scheme in both production processes, such terms which can be considered as an overall factor $(\mu_r^2)^\epsilon$ in each individual part (counter terms in the modified minimal subtraction ($\overline{\mathrm{MS}}$) scheme do not have such a factor), will vanish upon the cancellation of  UV and infrared (IR) divergences when summing over all the parts. This leads to $\mu_r$ independence of our results for the production processes. On the other hand, from Eq.~(\ref{eqn:scale}) a new scale $\mu$ is introduced, which denotes the scale at which charge is renormalized, and our results are dependent on it. From now on, when we talk about renormalization scale, it denotes $\mu$ which is introduced in the renormalization of charge, not $\mu_r$.

The renormalization of 2HDM had been plagued by the issue of gauge-dependence of the mixing parameters, like the $\tan \beta$, as in the MSSM case where type II 2HDM is needed \cite{Kanemura:2004mg,Kanemura:2015mxa,Kanemura:2017wtm}. A reasonable and convenient scheme should be gauge-independent and process-independent and numerically stable as well \cite{Freitas:2002um}.
Based on the renormalization scheme worked out by Fleischer and Jegerlehner  in Ref.~\cite{Fleischer:1980ub}, which is now usually called FJ tadpole scheme, two groups find a way to fulfill such conditions. The first one is the $\overline{\mathrm{MS}}$  tadpole scheme (MSTS)~\cite{Denner:2016etu,Denner:2017vms,Denner:2018opp}, where the mixing angles are renormalized using $\overline{\mathrm{MS}}$ subtraction. The other is the pinched tadpole scheme (PTS)~\cite{Krause:2016oke,Krause:2016xku}, where 
the pinch technique (see e.g. Ref.~\cite{Binosi:2009qm}) is used to define gauge-independent counter terms for the mixing angles. More renormalization schemes have been proposed and examined numerically \cite{Altenkamp:2017ldc,Altenkamp:2017kxk} and have been implemented in the HDecay package \cite{Krause:2018wmo}.
 
In this work, we adopt the on-shell PTS described in Ref.~\cite{Krause:2016oke}. On-shell here means that the on-shell scale is chosen in the self-energies during the renormalization of the mixing angles. More details about the renormalization scheme in this work can be found in Appendix \ref{appendixc}.

In our calculation, a small photon mass $\lambda$ is introduced to regularize the soft divergence. Meanwhile, two cutoffs, $\Delta E$ and $\Delta\theta$, are introduced to deal with the IR singularities in real correction processes. The 3-body phase space of real correction process $e^+e^-\rightarrow Zh\gamma$ is divided into three parts:
\begin{itemize}
\item soft ($S$) part: where the energy of photon $E_\gamma$ is smaller than $\Delta E$
\item hard collinear ($HC$) part: where $E_\gamma\ge\Delta E$ and the angle between photon and the beam $\theta_\gamma$ is smaller than $\Delta\theta$
\item hard non-collinear ($H\overline{C}$) part: the remaining, which is finite.
\end{itemize}
Thus next leading order (NLO) corrections can be expressed as
\begin{equation}
d\sigma^{1}=d\sigma_V(\lambda)+d\sigma_S(\lambda,\Delta E)+d\sigma_{HC+CT}(\Delta E,\Delta\theta)+d\sigma_{H\overline{C}}(\Delta E,\Delta\theta), \label{fullcor}
\end{equation}
Here $d \sigma_V$ denotes the virtual correction including loop diagrams and counter terms from renormalization.

The soft part is given by
\begin{eqnarray}
d\sigma_S&=&-\dfrac{\alpha_{em}}{\pi}d\sigma^0\times\biggl[\log\dfrac{4\Delta E^2}{\lambda^2}\biggl(1+\log\dfrac{m_e^2}{s}\biggr)+\dfrac{1}{2}\log^2\dfrac{m_e^2}{s}+\log\dfrac{m_e^2}{s}+\dfrac{1}{3}\pi^2\biggr],
\end{eqnarray}
where $d\sigma^0$ denotes the tree-level differential cross section. The hard collinear part is obtained in the collinear limit as
\begin{eqnarray}
d\sigma_{HC} =&&\dfrac{\alpha_{em}}{2\pi}\left[\dfrac{1+z^2}{1-z}\log\dfrac{\Delta\theta^2+4m_e^2/s}{4m_e^2/s}-\dfrac{2z}{1-z}\dfrac{\Delta\theta^2}{\Delta\theta^2+4m_e^2/s}\right]d\sigma^0(zk_1)dz +(k_1\Leftrightarrow k_2) \nonumber \\
\xrightarrow{\Delta\theta^2\gg m_e^2/s}&&\dfrac{\alpha_{em}}{2\pi}\left[\dfrac{1+z^2}{1-z}\log\dfrac{\Delta\theta^2s}{4m_e^2}-\dfrac{2z}{1-z}\right]\times\biggl[d\sigma^0(zk_1)+d\sigma^0(zk_2)\biggr]dz \,,
\label{eqn:HC}
\end{eqnarray}
with $0\le z \le 1-\delta_s=1-2\Delta E/\sqrt{s}$, which is also related to the tree-level differential cross section $d\sigma^0$. $CT$ denotes the ``counter term'' from electron structure function, originated from the 2nd term in Eq.~(\ref{eqn:ff}):
\begin{equation}
d\sigma_{CT}=-\dfrac{\alpha_{em}}{2\pi}\log\dfrac{s}{4m_e^2}P_{ee}^+(z,0)\times\biggl[d\sigma^0(zk_1)+d\sigma^0(zk_2)\biggr]dz
\end{equation}
with $0\le z \le 1$. Thus the combination of HC and CT parts can be expressed as
\begin{eqnarray}
d\sigma_{HC+CT}&\equiv& d\sigma_{HC+CT}^{*} + d\sigma_{SC} \nonumber \\
d\sigma_{HC+CT}^{*} &=&\dfrac{\alpha_{em}}{2\pi}\left[\dfrac{1+z^2}{1-z}\log\Delta\theta^2-\dfrac{2z}{1-z}\right]\times\biggl[d\sigma^0(zk_1)+d\sigma^0(zk_2)\biggr]dz\nonumber\\
d\sigma_{SC} &=&-\dfrac{\alpha_{em}}{\pi}\log\dfrac{s}{4m_e^2} \left[\dfrac{3}{2}+2\log\delta_s\right]d\sigma^0
\label{eqn:ISR}
\end{eqnarray}

Both the soft and virtual parts are obtained with FormCalc, while the other parts are obtained with the help of FDC~\cite{Wang:2004du}. The $\lambda$ dependence has been checked when we combine the soft and virtual parts, more detailed checking in the SM can be found in Appendix \ref{checkingpoints}.

In Appendix \ref{checkingpoints}, we examine the dependence of the SM cross-section on those unphysical parameters, such as $\delta_s$ in Table \ref{check:e} and $\Delta \theta$ in Table \ref{check:theta}. From these results, it is observed that within a reasonable region where $\delta_s\ll 1$ and $m_e/\sqrt{s} \ll \Delta\theta \ll 1$ are satisfied, our results are indeed independent of those unphysical parameters introduced by the algorithm. Meanwhile, as we adopt the $\alpha_{em}(m_Z)$ scheme, the cross-section should also be independent of the logarithm term of electron mass $\log(m_e)$. This independence is shown in Table \ref{check:me}. 
 After these checks, we apply our procedure to the 2HDM, i.e. we calculate the one-loop radiative corrections to the Higgs-strahlung process $e^+ e^- \to Z \phi$.

\begin{figure}[t!]\centering
\includegraphics[width=0.7\textwidth]{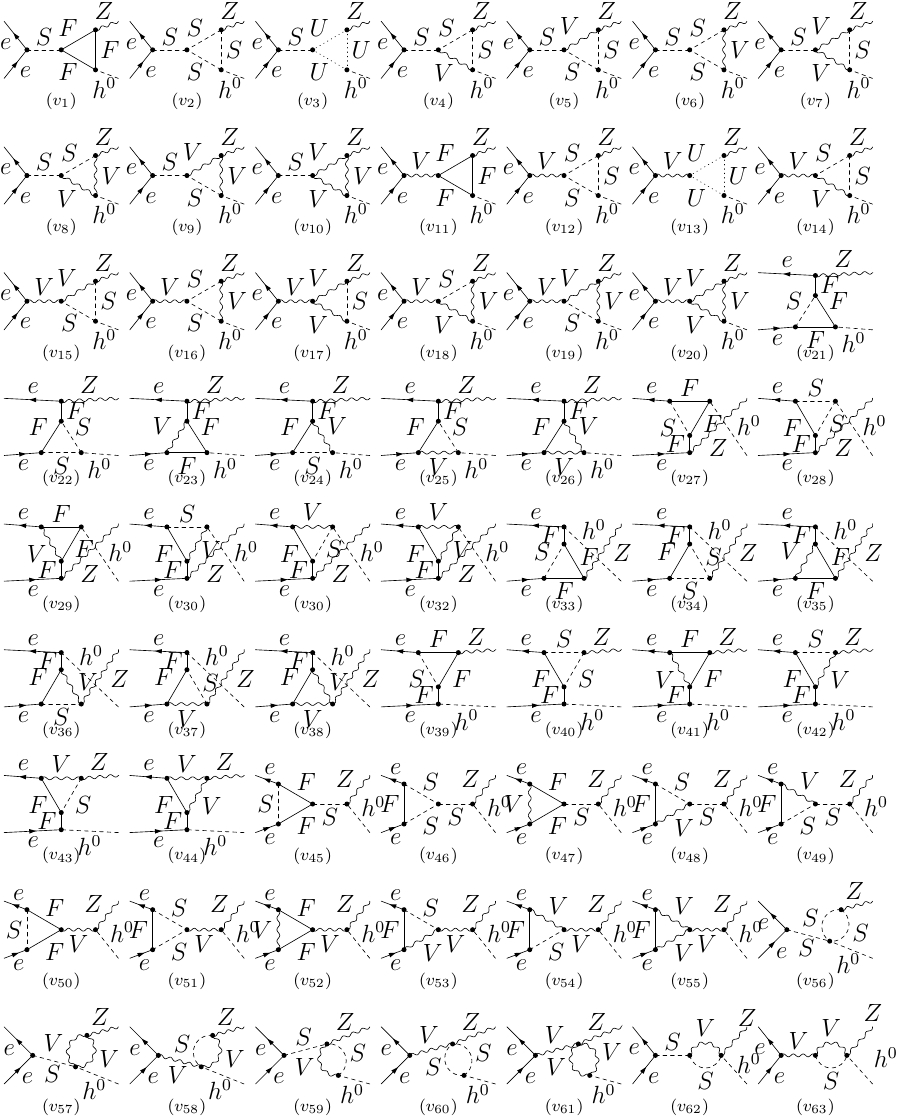}
\caption{Generic three-point one-loop Feynman diagrams in the 2HDM contributing to the process
$e^+ e^- \to Zh^0$ (we take $\phi = h^0$) are presented, where the labels $U$, $V$, $S$, and $F$ refer to ghosts, vector gauge bosons, Higgs scalar bosons, and fermions, respectively.}
\label{fig:vert}
\end{figure}
\begin{figure}[t!]\centering
\includegraphics[width=0.7\textwidth]{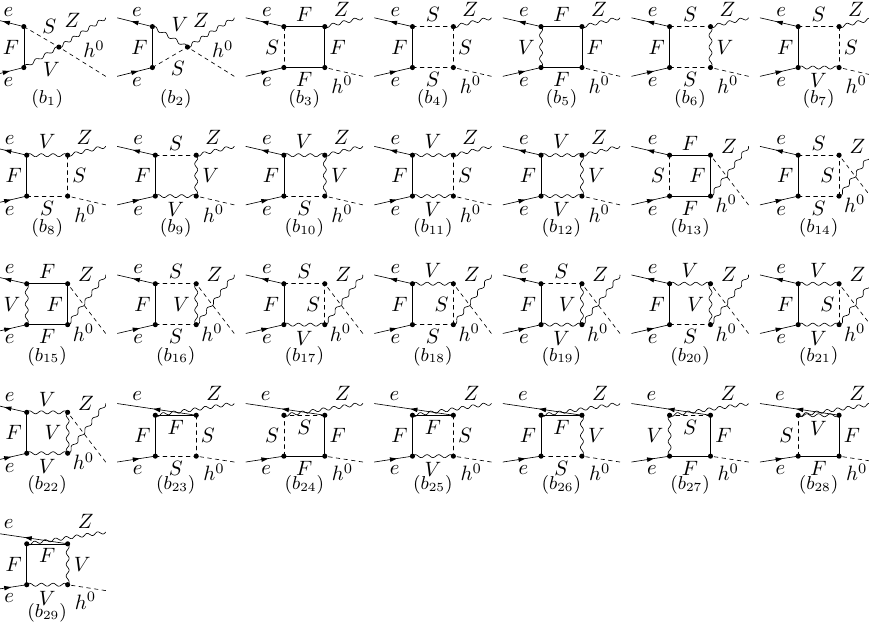}
\caption{Generic four-point one-loop Feynman diagrams in the 2HDM contributing to the process
$e^+ e^- \to Zh^0$ (we take $\phi = h^0$) are shown, where the labels of $V$, $S$,and $F$ refer to vector gauge bosons ,Higgs scalar bosons, and fermions, respectively. }
\label{fig:box}
\end{figure}
\begin{figure}[t!]\centering
\includegraphics[width=0.7\textwidth]{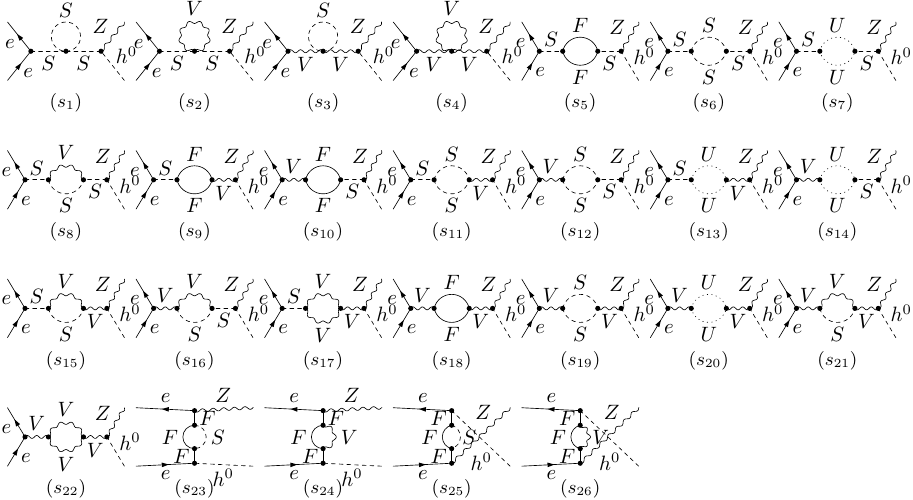}
\caption{Generic two-point one-loop Feynman diagrams in the 2HDM contributing to the process
$e^+ e^- \to Zh^0$ (we take $\phi = h^0$) are shown, where the labels of $U$, $V$, $S$, and $F$ refer to ghosts, vector gauge bosons, Higgs scalar bosons, and fermions, respectively.}
\label{fig:self}
\end{figure}
\begin{figure}[t!]\centering
\includegraphics[width=0.7\textwidth]{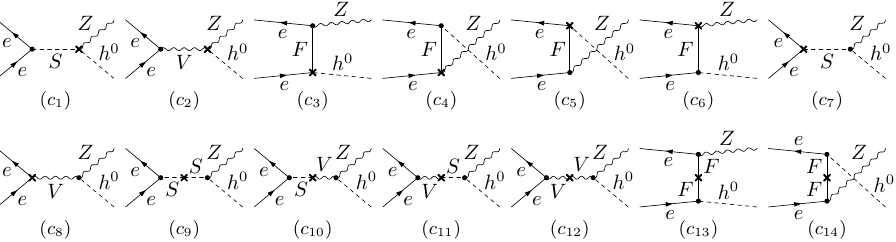}
\caption{Generic one-loop counter-terms in the 2HDM contributing to the process
$e^+ e^- \to Zh^0$ (we take $\phi = h^0$) are presented.}
\label{fig:ct}
\end{figure}

Here we calculate the radiative corrections to the tree level
$e^+ e^- \to Z h^0, Z H^0$ processes in 2HDM in the Feynman-'t Hooft gauge,
including all the particles of the model in the loops.
Counter-terms are constructed by specific renormalization conditions
which allow us to cancel all the UV divergences with one-loop diagrams. Inserting these redefinitions into the Lagrangian, we find the following counter term for $h^0 ZZ$ and $H^0 ZZ$:
\begin{eqnarray}
 \delta {\mathcal{L}}_{h^0 ZZ} &=&
\frac{e m_W \sin(\beta - \alpha)}{s_W c_W^2}
  \Big[\delta Z_e - \dfrac{\delta s_W}{s_W} -\dfrac{2\delta c_W}{c_W}+
\frac{\delta m_W^2}{2m_W^2} + \frac{1}{2} \delta Z_{h^0h^0} + \delta Z_{ZZ}
 \nonumber \\ &+& \frac{1}{2}\cot(\beta-\alpha)\delta Z_{H^0 h^0}  + \cot(\beta -\alpha)(\delta\beta-\delta\alpha) \Big]
h^0 Z^\mu Z^\nu g_{\mu\nu}\\
\delta {\mathcal{L}}_{H^0 ZZ} &=&
\frac{e m_W \cos(\beta - \alpha)}{s_W c_W^2}
  \Big[\delta Z_e - \dfrac{\delta s_W}{s_W} -\dfrac{2\delta c_W}{c_W}+
\frac{\delta m_W^2}{2m_W^2} + \frac{1}{2} \delta Z_{H^0H^0} + \delta Z_{ZZ}
\nonumber \\ &+& \frac{1}{2}\tan(\beta-\alpha)\delta Z_{h^0 H^0} - \tan(\beta -\alpha)(\delta\beta-\delta\alpha)\Big]
H^0 Z^\mu Z^\nu g_{\mu\nu} .
\end{eqnarray}

The one-loop Feynman diagrams related to the processes $e^+ e^- \to Z \phi_i$ are displayed in Figures (\ref{fig:vert}-\ref{fig:ct}),
which are conveniently dubbed as the vertex corrections, the box contributions, self-energy corrections and the counter terms, respectively. In these Feynman diagrams, the labels $V, S, F$ denote all insertions of vector, scalar and fermionic states of the 2HDM. 
As we used the on-shell PTS in our renormalization scheme, we should also include all possible tadpole diagrams, which have been merged to corresponding counter term diagrams in Figure (\ref{fig:ct}) and have not been shown explicitly.

There is one issue on the effects of the width of particles in the s-channel diagrams which need to clarify. For the processes that we consider here, the intermediate $Z$ boson is always far away from its mass shell as the collision energy  is assumed to be 250 GeV or even higher. Thus it is plausible to neglect the effect of the width of Z boson in our calculation. Meanwhile, it is observed that contributions from initial state $e^+e^-\phi$ vertices vanish in the limit $m_e \to 0$ since both $e^+$ and $e^-$ are on-shell, then it is justifiable to neglect the widths of scalar bosons. Obviously, such an argument holds for the rest of scalar boson exchange diagrams in s-channel.  As mentioned in the caption of Table \ref{BPdata}, the ratio $\Gamma/m$ for all the new Higgs boson particles is less than $5\%$ and the narrow width approximation can hold.  Therefore, in this work, for those s-channel diagrams, we simply neglect the widths of Z boson and other scalar bosons.

Finally, the total cross section at NLO, denoted as $\sigma^{NLO}$, is defined as the sum of LO cross section, $\sigma^0$, and the NLO corrections $\sigma^{1}$, i.e.
\begin{eqnarray}
\sigma^{NLO}&=&\sigma^{0} + \sigma^{1} \equiv \sigma^0(1+\delta^1)\, ,
\label{sig1}
\end{eqnarray}
where $\delta^1$ is defined as the ratio which measures the relative strength of next to leading order corrections over the tree-level result.

As described in Section 3.1 of Ref.\cite{Beenakker:1991ca}, the NLO electroweak corrections $\sigma^{1}$ can be safely grouped into two gauge-invariant parts:
\begin{itemize}
\item the "QED" part, which includes all the diagrams which contain an extra photon attached to the LO diagrams, such as diagrams $v_{23}$, $v_{29}$ and $v_{52}$ in Figure (\ref{fig:vert}) when the vector boson denotes a photon. Meanwhile, the photon's contribution to the wave-function renormalization of electron is also grouped into this part. 
\item the "weak" part, which contains all the remaining contributions.
\end{itemize}
In term of the convention introduced in Ref.\cite{Beenakker:1991ca}, we can divide $\sigma^1$ as
\begin{eqnarray}
\sigma^1 =\sigma^{\mbox{1,weak}}+\sigma^{\mbox{1,QED}}\,.\label{split0}
\end{eqnarray}
Correspondingly, the $\delta$ defined in Eq. (\ref{sig1}) can also be divided into two parts
\begin{eqnarray}
\delta^1 =\delta^{\mbox{1,weak}}+\delta^{\mbox{1,QED}}\,,\label{split}
\end{eqnarray}
which represent the relative strength of weak and QED corrections, respectively.

\section{Numerical results and discussions\label{sec:results}}
Below we present the numerical results that are obtained from the analysis of the processes
$e^+ e^- \rightarrow Z +h/H$ at the one-loop level in both the SM and the 2HDM. We focus on the following two quantities: the ratio of the weak correction $\delta^{\mbox{1,weak}}$ and the ratio of the full one-loop corrections (including real emissions) to the leading order results $\delta^1$.
\footnote{In the QED part, the extra photon can only attach to an initial electron pair in this process either in the SM or in the 2HDM. Thus $\delta^{\mbox{1,QED}}$ should be the same in both SM and 2HDM results. }

In our numerical calculation, parameters in Table~\ref{BPdata} are used as physical input in the Higgs sector, while in the gauge sector two different physical parameterizations are used:
\begin{itemize}
\item $\{\alpha_{em}, G_F, m_Z\}$: They are used in the scan of the 2HDM parameter space, which finally leads to Table~\ref{BPdata}. They are also used in determining the allowed range of $\lambda_5$ and in the calculation of decay part.
\item $\{\alpha_{em}, m_W, m_Z\}$: They are used in the calculation of production processes, as we have used on-shell condition for the renormalization of $W$ boson self-energy there.
\end{itemize}
The values of above parameters are taken from PDG~\cite{Tanabashi:2018oca} as $\alpha_{em}=1/137.036$, $G_F=1.16638\times 10^{-5}$ Gev$^{-2}$, $m_Z=91.1876$ GeV and $m_W=80.385$ GeV. It should be noted that the Vacuum Expectation Value (VEV) $v$ is different in these two cases. In the former case, it is determined as $v=1/\sqrt{\sqrt{2}G_F}\approx 246.220$ GeV. While in  the latter case, it is determined as $v=m_W\sqrt{1-m_W^2/m_Z^2}/\sqrt{\pi\alpha_{em}}$ and varies as the scale of $\alpha_{em}$ changes. In the $\alpha_{em}(m_Z)$ scheme, it is $v\approx 243.137$ GeV.

\begin{figure}[htbp]
\setcounter{subfigure}{0}
 \centering
 \subfigure[]
 {
 \includegraphics[width=0.45\textwidth]{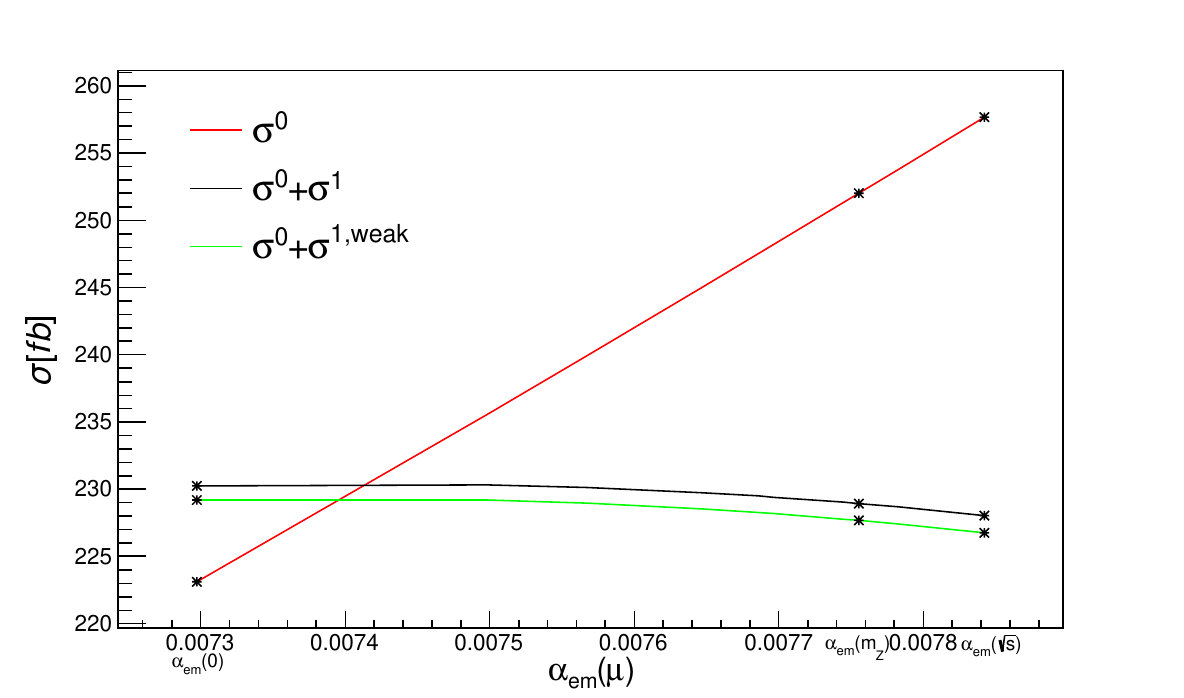}}
 \subfigure[]
 {
 \includegraphics[width=0.45\textwidth]{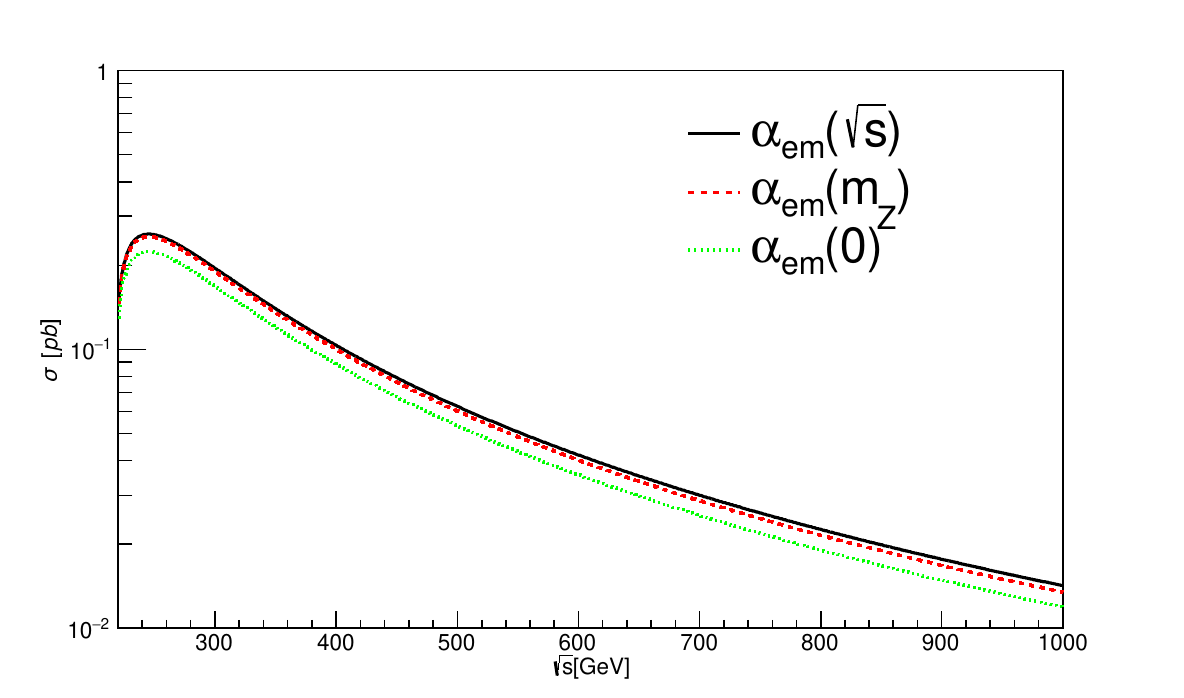}}
 \caption{(a): Scale dependence of LO and NLO cross section in the SM; (b): LO cross section as function of collision energy $\sqrt{s}$ in the SM in different schemes.} 
 \label{sigmu}
\end{figure}

The total NLO cross-section for $e^+e^-\rightarrow Z h_{SM}$ including the real emissions can be found in Table \ref{nlo_SM}. The results in the $\alpha_{em}(0)$ scheme still depend on $\log(m_e)$ while the other two are independent. At the LO, the cross-section $\sigma^0$ in these three schemes reads as 223.12 fb, 252 fb, and 257.68 fb, respectively, where the maximal difference is 34.56 fb. While at NLO, the cross-section $\sigma^{NLO}$ reads as 230.25, 228.93, and 228.05 fb, where the maximal difference is 1.20 fb. From these numbers, we can see that scale dependence has been greatly improved at NLO, which can be exposed more clearly in Figure  (\ref{sigmu}). For the sake of comparison with other literature, our definition of the weak part agrees to the one defined in Refs. \cite{Sun:2016bel,Gong:2016jys} and we also find that our numerical results in the SM agree with those given in Refs. \cite{Sun:2016bel,Gong:2016jys}.

{In Figure  (\ref{sigmu}),} the dependence on the renormalization scale $\mu$ and the collision energy of the cross-section in the SM are explicitly shown. In Figure  (\ref{sigmu}a), it is observed that the leading order results can change drastically when the unphysical renormalization scale $\mu$ varies from 0 to $\sqrt{s}$, i.e. the difference between the results of $\alpha_{em}(0)$ and $\alpha_{em}(\sqrt{s})$ can reach up to $15\%$ as given in Table \ref{nlo_SM}. In contrast, the scale dependence of the results is significantly reduced at the next-leading order. For example, the difference between the NLO results of $\alpha_{em}(0)$ and $\alpha_{em}(\sqrt{s})$ is $1\%$ or so. In Figure  (\ref{sigmu}b), the lineshapes of the LO and NLO cross sections varying with the collision energy $\sqrt{s}$ are shown. The largest difference between different renormalization scales occurs near the threshold.  

\begin{figure}[htbp]
\setcounter{subfigure}{0}
 \centering
 \subfigure[]
 {
 \includegraphics[width=0.45\textwidth]{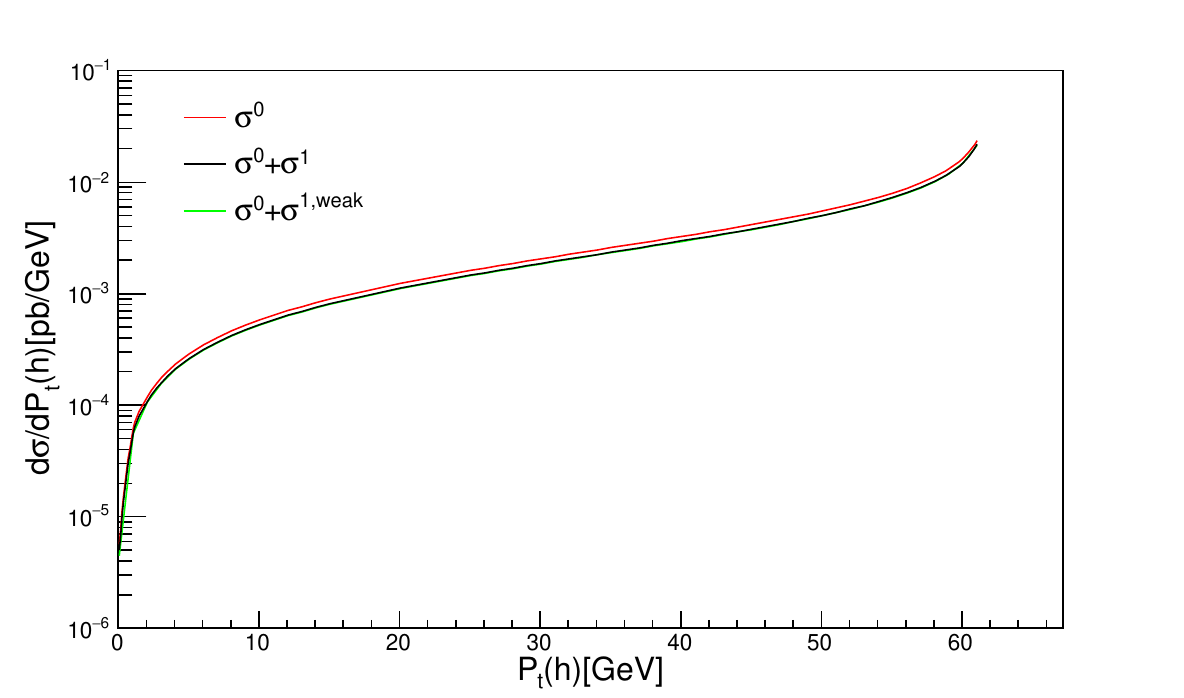}}
 \subfigure[]
 {
 \includegraphics[width=0.45\textwidth]{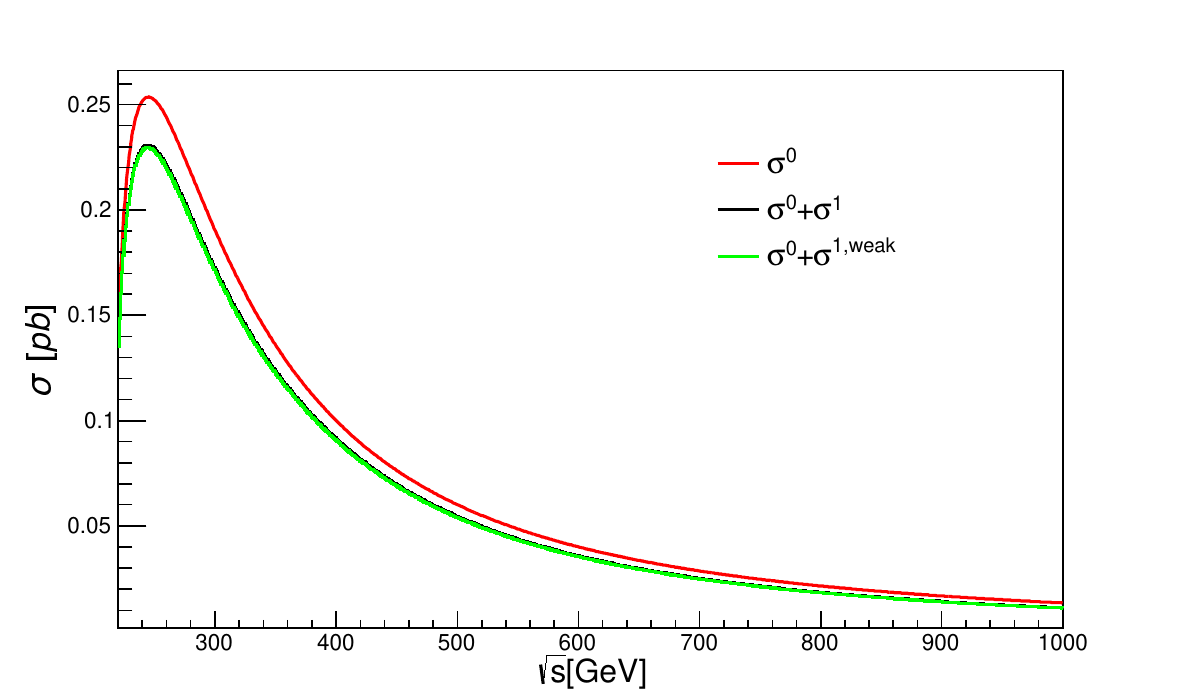}}
 \caption{(a): Transverse momentum $P_t(h)$ distribution of LO and NLO cross section; (b): LO and NLO cross sections as functions of collision energy $\sqrt{s}$ in the SM in the $\alpha_{em}(m_Z)$ scheme. } \label{nlo}
\end{figure}

In Figure  (\ref{nlo}), we compare the results of leading order and those of the full NLO in the SM.
In Figure  (\ref{nlo}a), the distribution of the transverse momentum of the Higgs boson in the process $e^+ e^- \to Z h $ in the SM is shown where the collision energy $\sqrt{s}$ is taken as 250 GeV. When the angle of the outgoing Higgs boson with reference to the incoming electron direction is defined as $\theta$ in the laboratory frame, the transverse momentum of the Higgs boson is defined as $P_t(h)= |\overrightarrow{P}(h)| \sin \theta$, where $|\overrightarrow{P}(h)|$ denotes the magnitude of 3-dimensional momentum of Higgs boson in the laboratory frame. The cutoff of $P_t(h)$ is determined by the total collision energy. When the collision is set to be 250 GeV, the maximum of $P_t(h)$ is $62.12$ GeV which is determined by kinematics. We find that the line-shapes of the LO and full next-to-leading order (NLO) results are similar except a global scaling factor. In Figure  (\ref{nlo}b), the energy dependence of the cross-section is shown.

 As a cross check, we compare our LO cross section of the SM with that computed using Whizard \cite{Kilian:2007gr} (say in the $\alpha_{em}(m_Z)$ scheme), and a good agreement is found. Moreover, it is observed that although the lineshape of $P_t(h)$ given in Figure (\ref{nlo}a) is different from Figure (2b) in \cite{Chen:2017gzv} where the distribution is a normalized one with the effects of initial state radiation (ISR) included, a good agreement is found when we also normalize the lineshape of $P_t(h)$ given in Figure  (\ref{nlo}a) with the tree-level results of Figure  (2b) of \cite{Chen:2017gzv} simulated using MadGraph\cite{Alwall:2014hca}.

\begin{table}[http]
\begin{center}
\begin{tabular}{|c|c|c|c|c|c|c|}
\hline\hline
scheme&$1/{\alpha_{em}(\mu)}$&$\sigma^{0}$&$\sigma^{1,weak}$&$\sigma^{1}$&$\sigma^{NLO}=\sigma^{0}+\sigma^{1}$\\
\hline
$\alpha_{em}(0)$     &137.036&223.12(0)&  6.09(0)&  7.13(2)&230.25(2)\\
$\alpha_{em}(m_Z)  $&128.943&252.00(0)&-24.33(0)&-23.07(2)&228.93(2)\\
$\alpha_{em}(\sqrt{s})$&127.515&257.68(0)&-30.92(0)&-29.63(2)&228.05(2)\\
 \hline
\end{tabular}
\caption{NLO SM results in different schemes at $\sqrt{s} = 250$ GeV are shown (in unit of fb). Three schemes are chosen to demonstrate the scale dependence of the results.}
\label{nlo_SM}
\end{center}
\end{table}

\begin{figure}[htbp]
\setcounter{subfigure}{0}
 \centering
 \subfigure[BP1-h]{
 \includegraphics[width=0.45\textwidth]{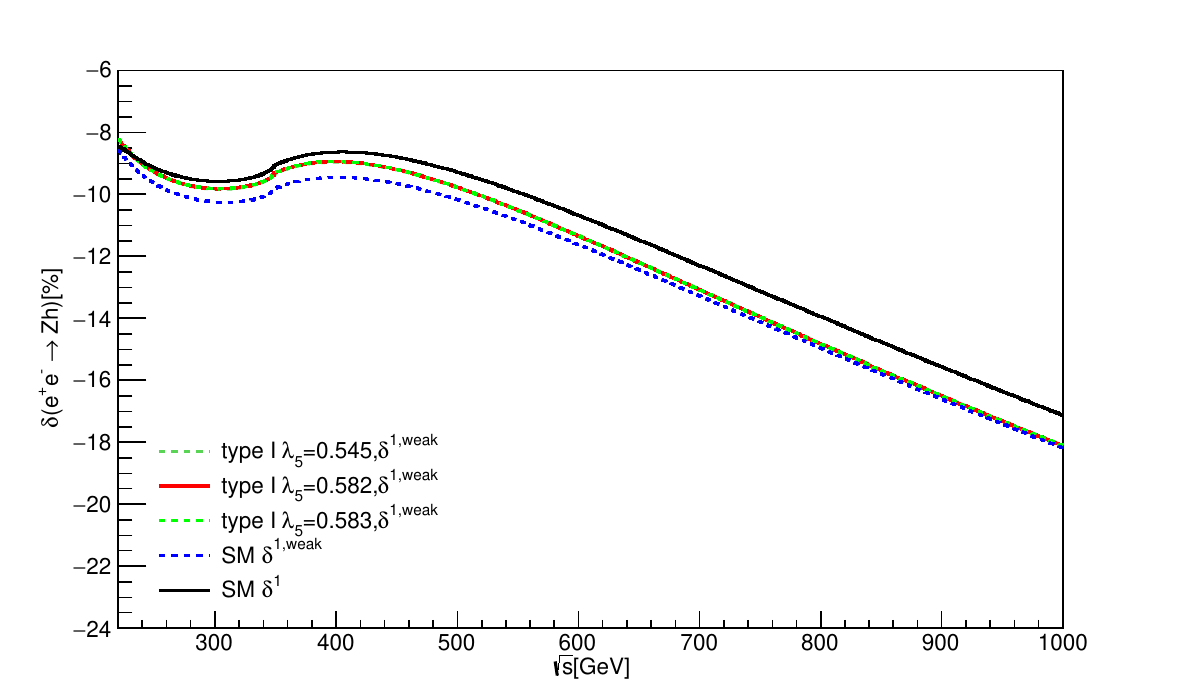}}
 \subfigure[BP2-h]{
 \includegraphics[width=0.45\textwidth]{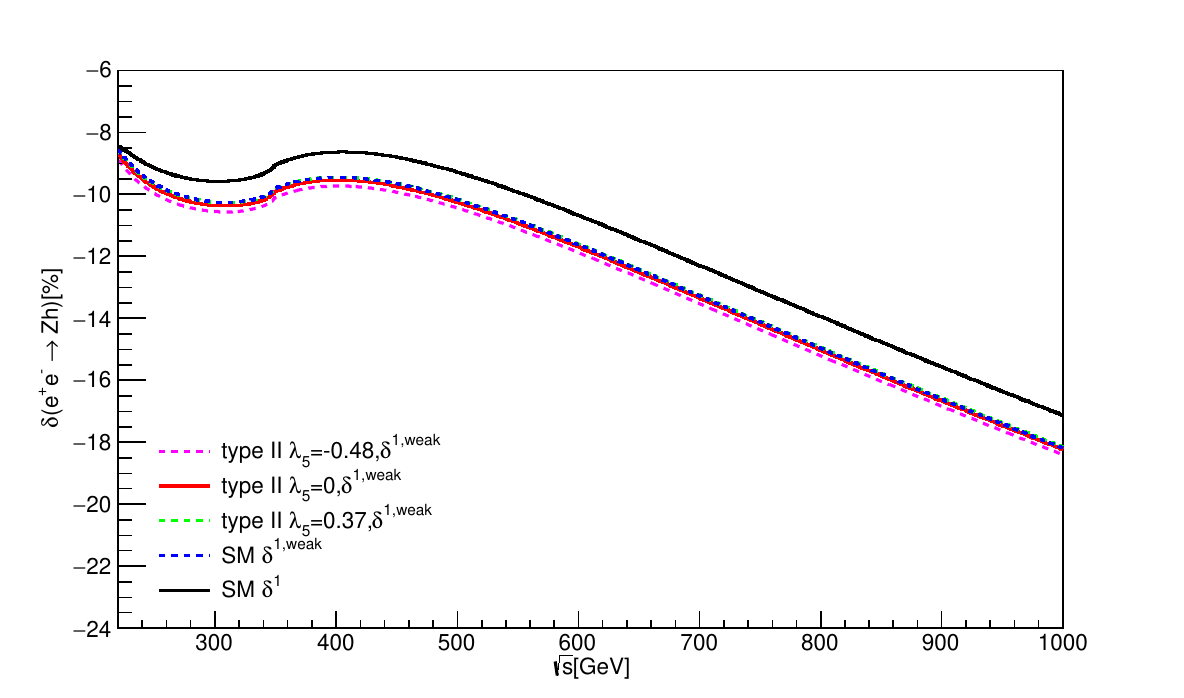}}
 \caption{Ratios of weak and full corrections to LO results for $e^+ e^-\rightarrow Zh$ as functions of collision energy corresponding to benchmark points BP1-h (left) and BP2-h (right) in the 2HDM, where $h^0$ is assumed to be the SM-like Higgs boson. A few typical values of $\lambda_5$ are taken to show the effects of triple Higgs couplings. }
\label{fig1:2hdm2}
\end{figure}

Based on the results of the SM, since the NLO results have less dependence on the renormalization scale $\mu$, in the following study on the 2HDM, we will fix a renormalization scale $\mu=m_Z$ to present the results.

Now we start to present the results of the four benchmark points of 2HDM. 
The parameter of $\lambda_5$ is related to the triple Higgs boson couplings, which is given as \cite{Gunion:2002zf}
\begin{equation}
\lambda_5 = \frac{m^2_{12} - m^2_A s_\beta c_\beta}{v^2 s_\beta c_\beta} \label{l5m122}
\end{equation}
Once $\lambda_5$ is fixed, all the triple Higgs boson couplings are fixed, as shown in Eqs. (\ref{lll}-\ref{hpH}) of Appendix \ref{trihiggsc}.
To demonstrate the effects of triple Higgs boson self-couplings, we try to vary it near its value for each benchmark point in Table \ref{BPdata}.
From now on, we treat each BP as a scenario.
In each scenario, we fix the mass spectra, $\alpha$ and $\tan\beta$ as given in Table II, and let $\lambda_5$ vary.
Due to the constraints from vacuum stability, unitarity and perturbativity to the theoretical parameters, the parameters $\lambda_{5}$ for 4 benchmark scenarios are determined to be in the ranges [0.545,0.583], [-0.48, 0.37], [-1.32,-0.30] and [-6.50,-5.74], respectively. We present our results in 2HDM with three typical values of $\lambda_5$, the upper bound, lower bound and the one in the benchmark point.
 
In Figure (\ref{fig1:2hdm2}), the ratios $\delta^{1,weak}$ and $\delta^{1}$, which defined in Eq. (\ref{split}), are shown, 
where the results are obtained in the $\alpha_{em}(m_Z)$ scheme and label "$h$" denotes that the lighter CP-even Higgs boson is assumed to be the SM-like Higgs boson. $\delta^{1}$ in which both weak and QED corrections are included is only presented for the SM, as we have discussed before that $\delta^{1,QED}$ is exactly the same in both SM and 2HDM. 
It is found that the signs of weak part and QED part are different in the SM. The QED part tends to increase the total cross-section by $+0.5\sim+1\%$ as the collision energy increases.

Also, in the BP1-h and BP2-h scenarios, the difference among the curves with different values of $\lambda_5$ is small and insignificant. For BP1-h scenario, this may be attributed to the fact that the allowed range for $\lambda_5$ is too narrow and the triple Higgs boson couplings are small, as shown in Figure (\ref{figthc:thcs}a). In contrast, for the BP2-h scenario, the allowed range is wide enough and the triple Higgs couplings can be large enough, as shown in Figure (\ref{figthc:thcs}b), this small difference instead can be attributed to the decoupling effects of heavy Higgs boson. It is found that for BP2-h scenario, this difference between the prediction of the SM is small and typically less than $0.2\%$ at most, which means at NLO new physics plays a less important role. While for BP1-h scenario, it it observed that the difference between the new physics and the SM can reach $0.8\%$ or so from 300 GeV to 500 GeV.

%

\begin{figure}[htbp]
\setcounter{subfigure}{0}
 \centering
 \subfigure[BP1-H]{
 \includegraphics[width=0.45\textwidth]{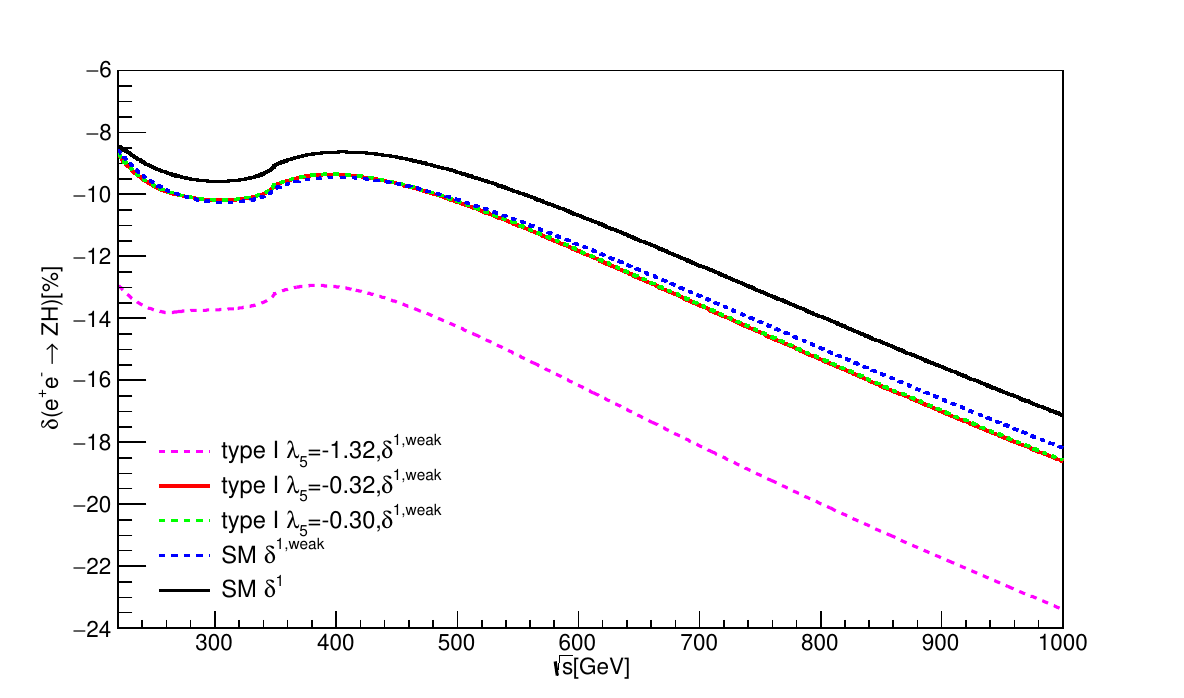}}
 \subfigure[BP2-H]{
 \includegraphics[width=0.45\textwidth]{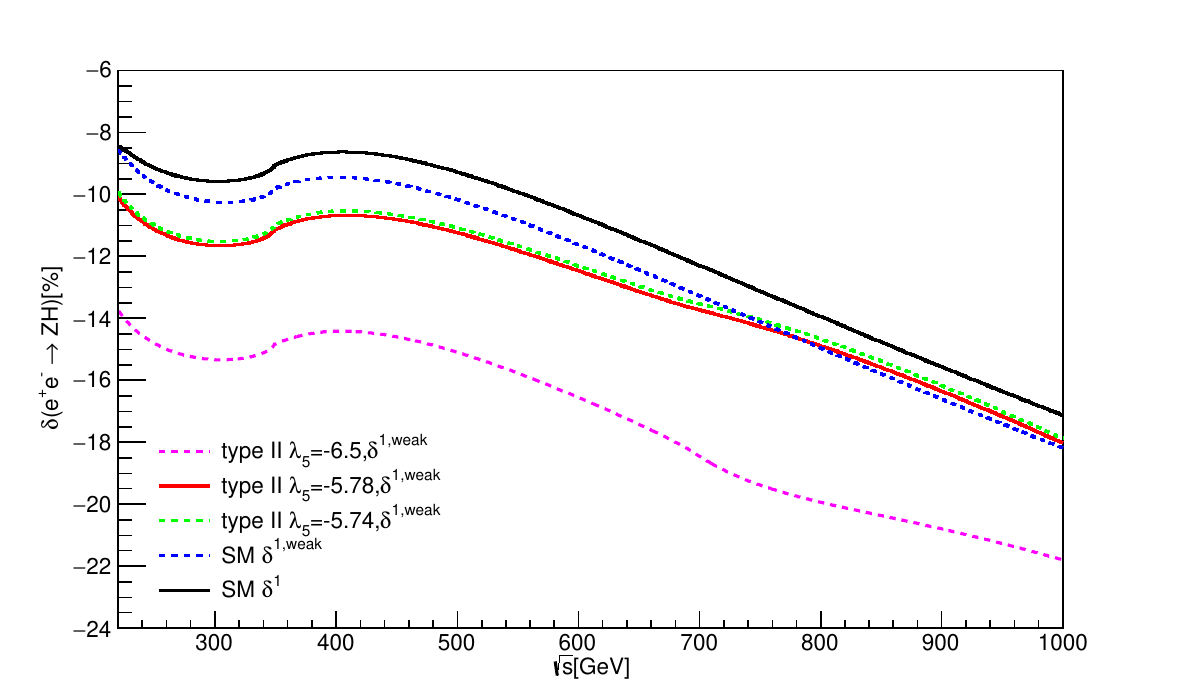}}
 \caption{Ratios of weak and full corrections to LO results for $e^+ e^-\rightarrow ZH$ as functions of collision energy corresponding to benchmark points BP1-H (left) and BP2-H (right) in the 2HDM. A few typical values of $\lambda_5$ are taken to show the effects of triple Higgs couplings. }
\label{fig2:2hdm4}
\end{figure}

In Figure  (\ref{fig2:2hdm4}), two benchmark points for an alternative interpretation of the SM-like Higgs boson are considered, in such a scenario the heavier CP even Higgs boson has a mass near 125 GeV. The detailed information on the mass spectra and parameters of these two benchmark points are presented in Table \ref{BPdata} and labeled as BP1-H and BP2-H scenarios. 
Again, three typical values of $\lambda_5$ are used to show the effects of new physics.
 We observe that the ratio for BP1-H scenario  (the pink dashed curve) varies from $-13\%$ to $-23\%$ or so, while for BP2-H scenario it varies from $-13.8\%$ to $-22\%$. Also in this alternative interpretation, the ratio difference for different values of $\lambda_5$ (say $\lambda_5=-1.32$ and $\lambda_5=0.30$ for BP1-H scenario, and $\lambda_5=-6.5$ and $\lambda_5=-5.74$ for BP2-H scenario) can reach  more than $5\%$.
It is found that the weak corrections of 2HDM for these two benchmark points and the typical values of $\lambda_5$ have the same sign in the SM, which holds for both the type-I and type-II as well as for the process $e^+ e^- \to Z h$ and $e^+ e^- \to Z H$. The last but not the least, all the lines have a bump near the collision energy 350-400 GeV, which can be attributed to the contribution of top quark pair in the loop functions.

\begin{figure}[htbp]
\setcounter{subfigure}{0}
 \centering
 \subfigure[BP1-h]{
 \includegraphics[width=0.45\textwidth]{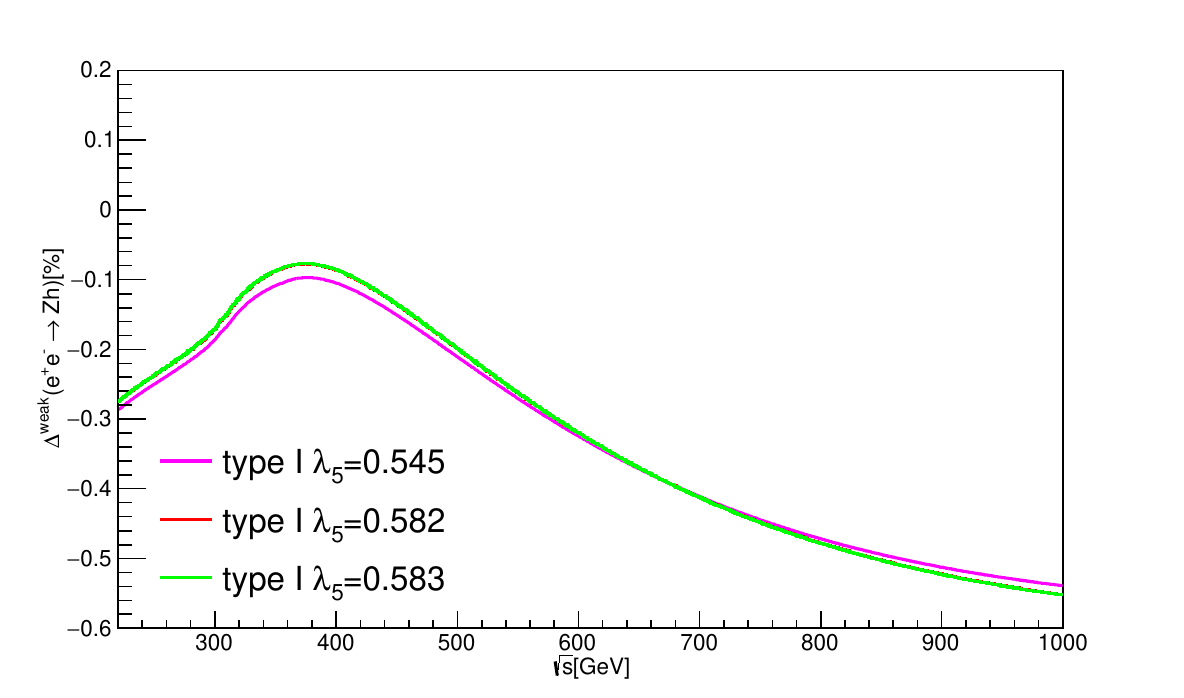}}
 \subfigure[BP2-h]{
 \includegraphics[width=0.45\textwidth]{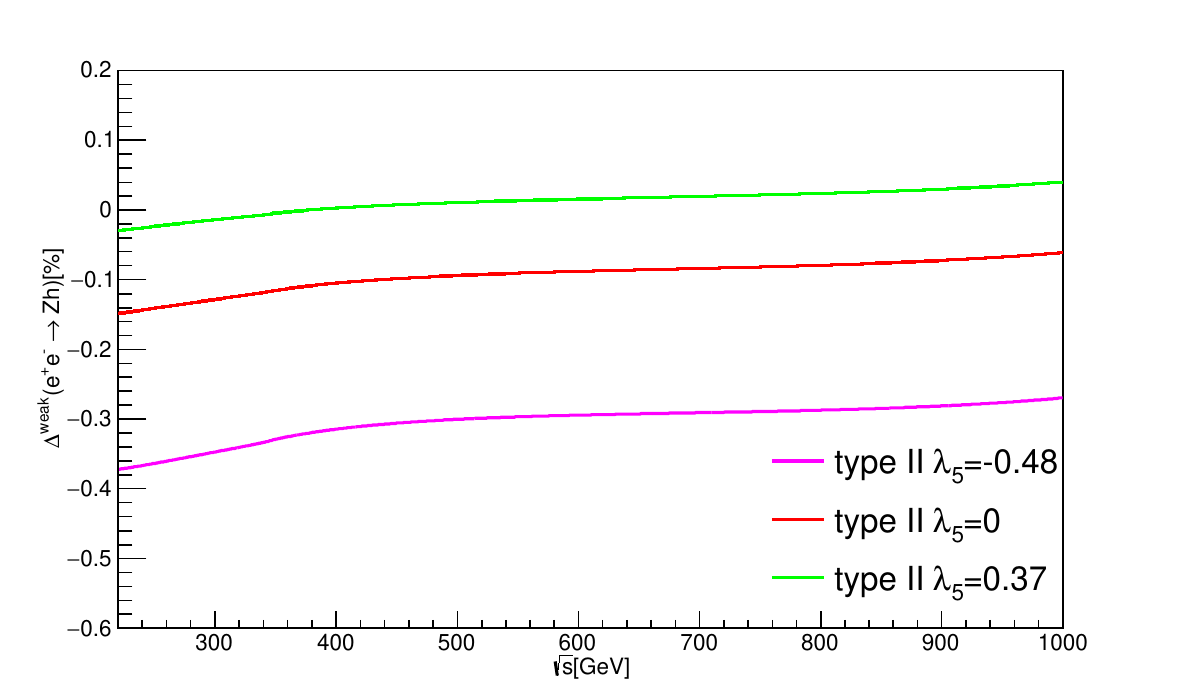}}
 \caption{Ratios of new physics for process $e^+ e^-\rightarrow Zh$ as functions of collision energy with three typical values of $\lambda_5$}
\label{fig1:weakthdm3}
\end{figure}

\begin{figure}[htbp]
\setcounter{subfigure}{0}
 \centering
 \subfigure[BP1-H]{
 \includegraphics[width=0.45\textwidth]{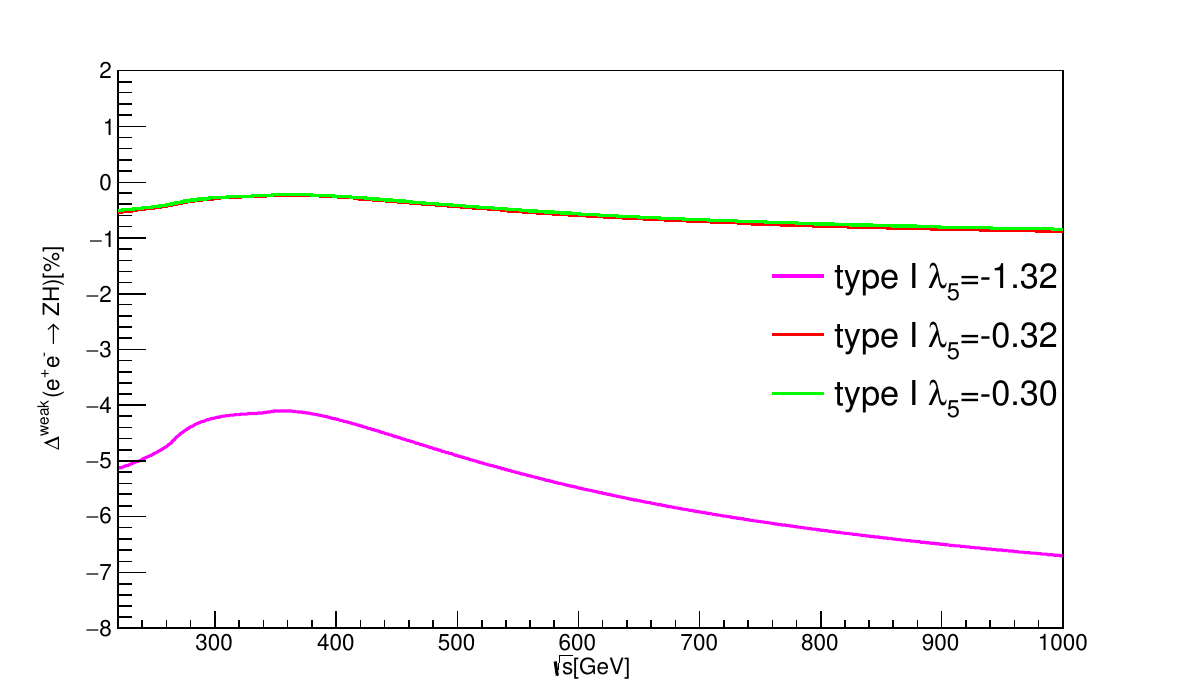}}
 \subfigure[BP2-H]{
 \includegraphics[width=0.45\textwidth]{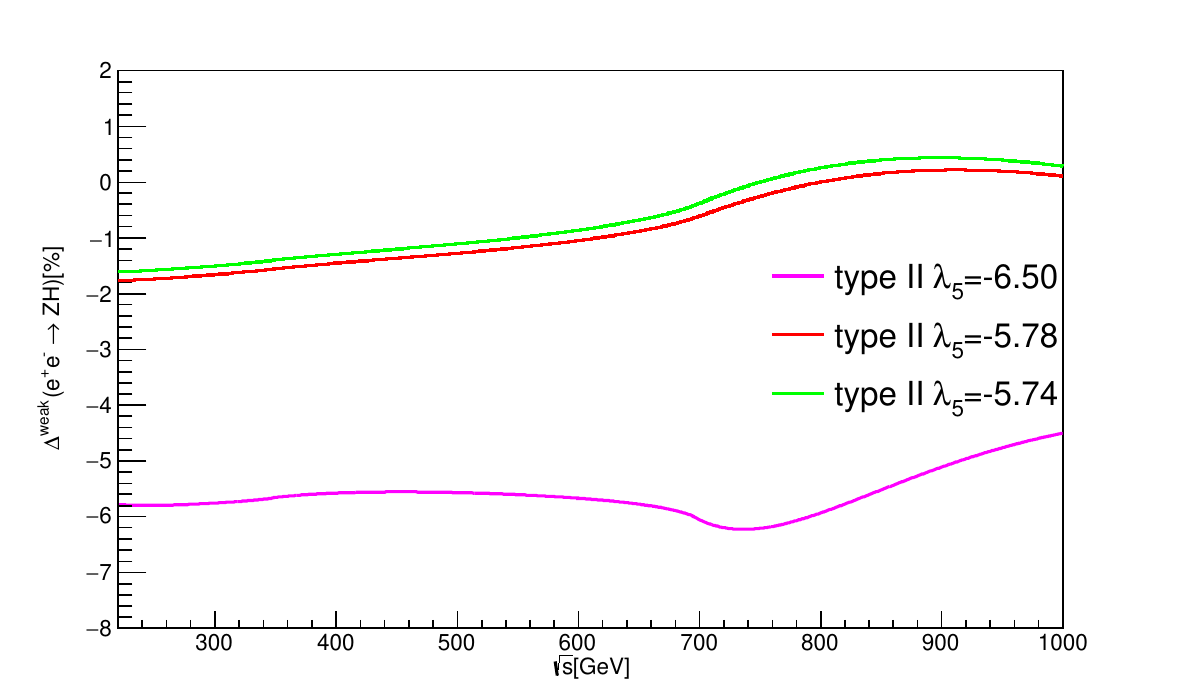}}
 \caption{Ratios of new physics for process $e^+ e^-\rightarrow ZH$ as functions of collision energy with three typical values of $\lambda_5$}
\label{fig2:weakthdm4}
\end{figure}

In order to expose the contribution of new physics, in Figures  (\ref{fig1:weakthdm3}) and (\ref{fig2:weakthdm4}), we present the contribution of new physics $\Delta^{weak}_{Zh}$ and $\Delta^{weak}_{ZH}$, respectively with different values $\lambda_5$. 
The quantity $\Delta^{weak}$ is defined as
\begin{equation}
\Delta^{weak} = \frac{\sigma_{2HDM}^0 + \sigma_{2HDM}^{1,weak}}{\sigma_{SM}^0 + \sigma_{SM}^{1,weak}} - 1 ,
\end{equation}
which describes the contribution of new physics compared with the results of the SM. We have discussed before that $\delta^{1,QED}$ are exactly same in both the SM and 2HDM, thus $\Delta^{weak}$ is enough to show the differences.
Obviously, in this formula, the pure contribution of the SM is subtracted, while the contribution of new physics and the interference terms between the new physics and the SM are counted.

 In Figure (\ref{fig1:weakthdm3}), the quantities $\Delta^{weak}$ of benchmark scenarios BP1-h and BP2-h are demonstrated. From the figure in the left panel, it is observed that the deviations caused by the 2HDM can change from $-0.28\%$ to $-0.29\%$ near the threshold and can decrease to $-0.54\%$ and $-0.52\%$ when the collision energy increases to 1000 GeV. A bump near the $2 m_{H^\pm} = 356$ GeV can be seen for BP1-h scenario.  From the figure in the right panel, the deviations are flat in the explored energy region and can change from $-0.03\%$ to $-0.37\%$ at the threshold and can decrease to $+0.02\%$ and $-0.30\%$ at 1 TeV. Meanwhile, the deviations are more sensitive to $\lambda_5$ in BP1-h scenario where new Higgs bosons are relatively lighter than those in BP2-h scenario, although the allowed range of $\lambda_5$ is narrower. For three cases of BP2-h scenario, the deviations show no obvious bumps within the explored collision energy.

 In Figure (\ref{fig2:weakthdm4}), the quantities $\Delta^{weak}$ of the benchmark scenarios BP1-H and BP2-H are shown. In the figure of left panel for the BP1-H scenario, the deviations are around $-0.58\%$ at the threshold and are similar for both $\lambda_5=-0.32$ and $\lambda_5=-0.30$. For two cases with $\lambda_5=-0.32$ and $\lambda_5=-0.30$, when the collision energy increases to 1000 GeV, the deviations change to $-1.5\%$ or so. While the case with $\lambda_5=-1.32$, the deviation changes from $-5.2\%$ at the threshold to $-6.6\%$ at 1000 GeV. In the figure of right panel for the BP2-H scenario, the deviations can reach from $-1.6\%$ to $-5.8\%$ which depends upon the values of $\lambda_5$. When the collision energy increases to 1000 GeV, the deviations change to $0.1\%$ to $-4.4\%$.

For the case of BP1-H scenario with $\lambda_5=-1.32$, a little bump near the energy region $\sqrt{s} = 2 m_{H^\pm}=340$ GeV can be observed. In contrast, for the two cases of BP1-H scenario with $\lambda_5=-0.30$ and $\lambda_5=-0.32$, the new physics contributions show no clear structure with the increase of collision energy due to smaller triple Higgs couplings. For the BP2-H case with $\lambda_5=-6.50$, an obvious dip near the energy region $\sqrt{s}=m_{A^0}+m_{h^0}=695$ GeV can be attributed to the contribution of $A^0$ and $H^0$ via the diagrams like $v_{12}$, $v_{14}$, $v_{15}$ and $v_{16}$, in Figure (\ref{fig:vert}).

Except for the differential cross-sections of $e^+ e^- \to Z \phi_i$ for four benchmark points in the 2HDM, it might also be interesting to explore the effects of new physics to the branching fractions of Higgs boson decays, which can be measured to a precision of $1\%$ or less at future Higgs factories.\\

The decay width of neutral Higgs boson in a model including one-loop correction can be formulated as 
\begin{eqnarray}
\Gamma_1 (\phi \to f {\bar f}) = \frac{N_C  G_F m_f^2}{4 \sqrt{2} \pi} \beta_f^3 (\xi^{\phi}_f)^2 m_\phi \widehat{Z}_{\phi} [1 - \Delta r + 2 \mathrm{Re}(\Delta {\cal M}_1)]\,,
\label{decayw}
\end{eqnarray}
where $N_C$ is number of color, $G_F$ is the Fermion constant, $m_f$ denotes the fermion's mass, $\beta_f$ is the speed of fermion in natural units, and $\xi^{\phi}_f$ is the ratio of couplings of neutral Higgs boson in the model over those of the SM. The $ \widehat{Z}_{\phi}$ refers to the finite wave function renormalization of the external $\phi$ while $\Delta{\cal M}_{1}$ is the amplitude of the one-loop vertex diagrams of $\phi\to f\bar f$. The quantity $\Delta r$ can be found in \cite{Denner:1991kt}, which is obviously model dependent and renormalization scheme dependent. 
As this part is a work following Ref.~\cite{Arhrib:2004ak}, the same renormalization scheme as in Ref.~\cite{Arhrib:2004ak} is used. 
Our results for the decays are renormalization scale dependent since the $\overline{\mathrm{MS}}$ scheme is used in the renormalization of Higgs boson wave functions. We have chosen the scale as $\mu_r=m_\phi$, the mass of decaying particle.
For the study on $\phi \to f\bar{f} $ decay, as it is a work following Ref.~\cite{Arhrib:2004ak}, the same renormalization scheme as in Ref.~\cite{Arhrib:2004ak} is used and will not be discussed here.

We can define a quantity $\Delta_{ff} (\phi)$ from Eq. (\ref{decayw})  as a deviation from the predictions of the SM, which has the following form
\begin{eqnarray}
 \Delta_{ff} (\phi)&=&\frac{ \widehat{Z}_{\phi}(1 -\Delta r^{2HDM}+
2 \mathrm{Re} (\Delta {\cal M}_1^{2HDM}) ) }{(1 -\Delta r^{SM}+
2 \mathrm{Re} (\Delta {\cal M}_1^{SM}) ) } -1, \ \ f=b, \tau \, .
\label{eq:deltaff}
\end{eqnarray}
More details can be found in \cite{Arhrib:2016snv}.

It is necessary to point out that the value of $\Delta_{ff} (\phi)$ in the SM is $0$, and it is a measurement of new physics effects at quantum level. 
There are several studies which evaluate the BSM effects at the tree-level and one-loop radiative corrections to the decays of Higgs bosons \cite{Arhrib:2004ak} and recently in \cite{Kanemura:2019slf, Krause:2019qwe}. 

In the type-I and type-II models, the couplings of $h^0$ to $b\bar{b}$ and $\tau^+ \tau^-$ are proportional to the same factor $\xi^{h^0}_f$. 
It is $\cos\alpha/\sin\beta$ in type-I and $-\sin\alpha/\cos\beta$ in type-II.
The couplings of $H^0$ to $b\bar{b}$ and $\tau^+ \tau^-$ are also proportional to another same factor $\xi^{H^0}_f$. In type-I  it is $\sin\alpha/\sin\beta$ while in type-II, it is $\cos\alpha/\cos\beta$. Obviously, there exist two special cases: 1) the couplings $|\xi^{h^0}_f| = 1$ if $h^0$ corresponds to the SM-like Higgs boson; 2) the couplings  $|\xi^{H^0}_f| = 1$ if $H^0$ corresponds to the SM-like Higgs boson. Apparently, from tree-level expression, it is impossible to separate these cases from the SM. Below we explore whether it is possible to distinguish these cases by using the quantum corrections, i.e. by resorting to the information of $\Delta_{ff}(h^0)$ and $\Delta_{ff}(H^0)$.




\begin{figure}[h!]
\begin{center}
\includegraphics[width=0.45\textwidth]{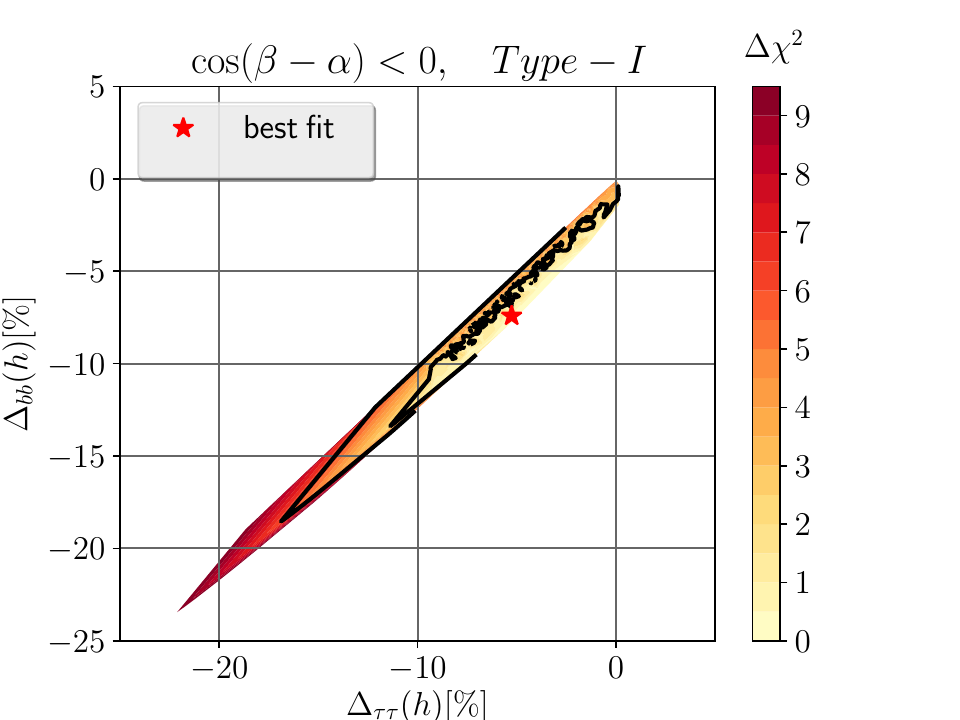}
\includegraphics[width=0.45\textwidth]{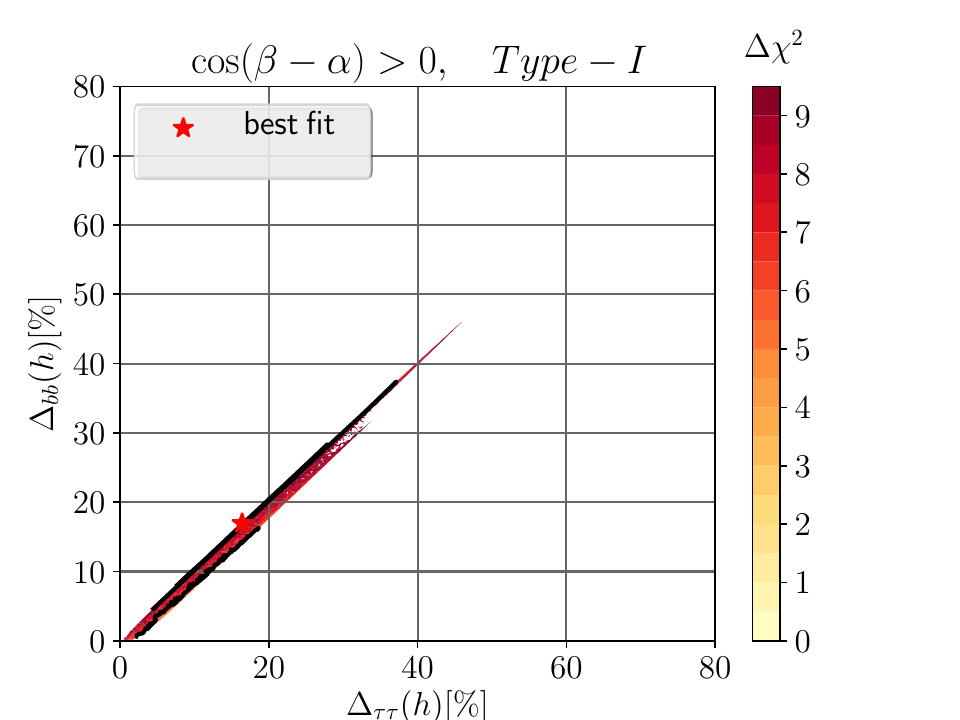}
\includegraphics[width=0.45\textwidth]{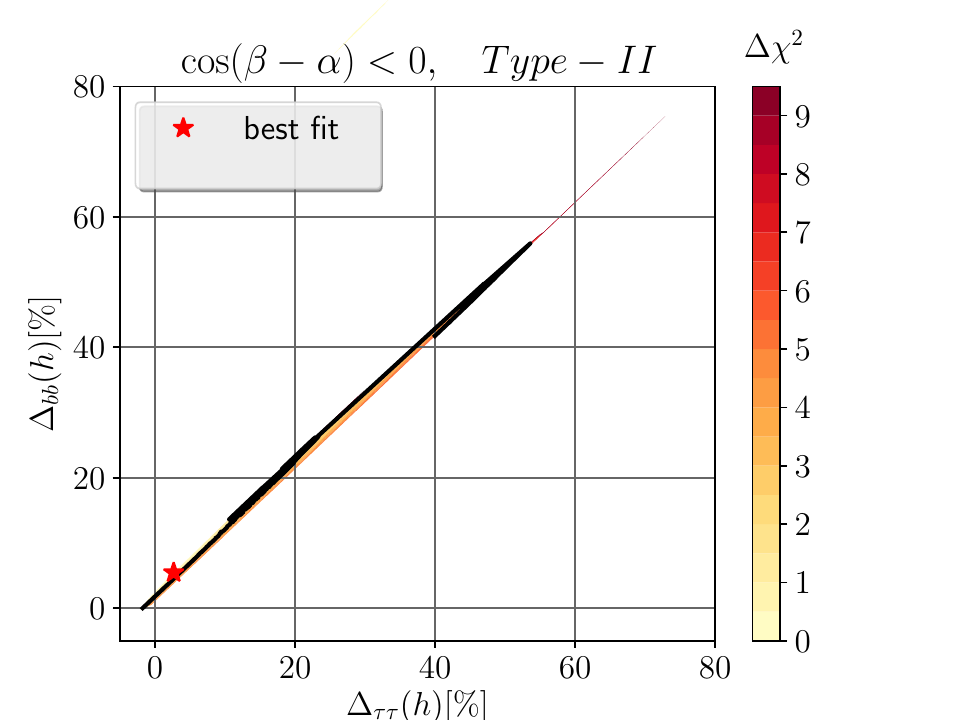}
\includegraphics[width=0.45\textwidth]{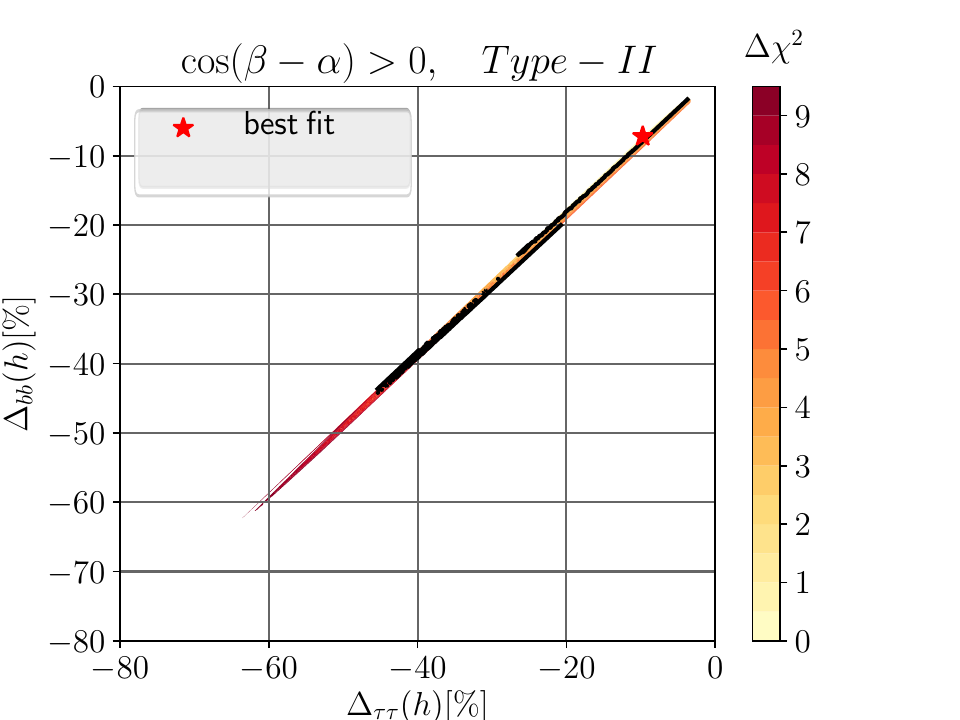}
\end{center}
\caption{Correlation between relative precisions $\Delta_{\tau\tau}(h)$ and
 $\Delta_{bb}(h)$ are examined for the 2HDM, where $h^0$ is assumed to be the SM-like Higgs boson with $m_H$ = 212 GeV (594 GeV) for the $\pm$D1-h scenarios (upper panel) and the $\pm$D2-h scenarios (lower panel). }
\label{fig:best-fit}
\end{figure}

\begin{figure}[h!]
\begin{center}
\includegraphics[width=0.45\textwidth]{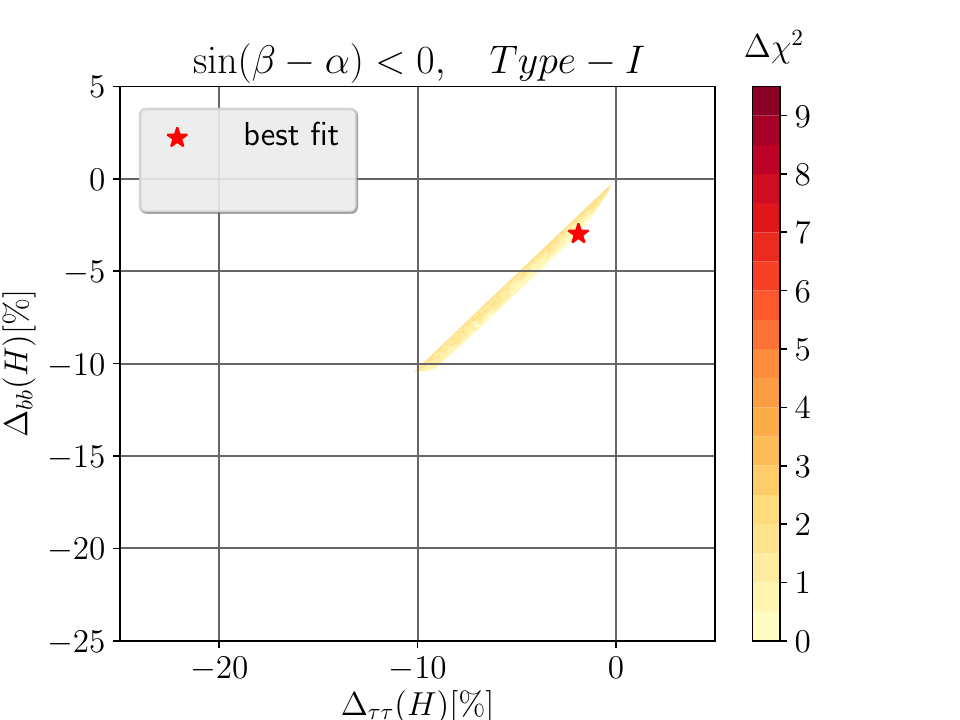}
\includegraphics[width=0.45\textwidth]{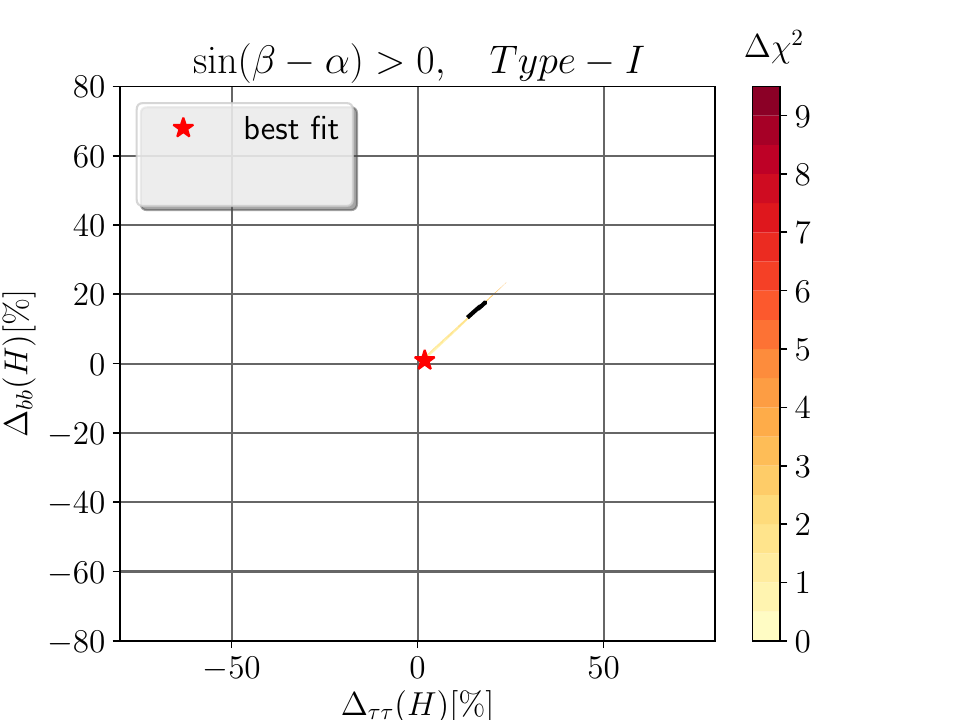}
\includegraphics[width=0.45\textwidth]{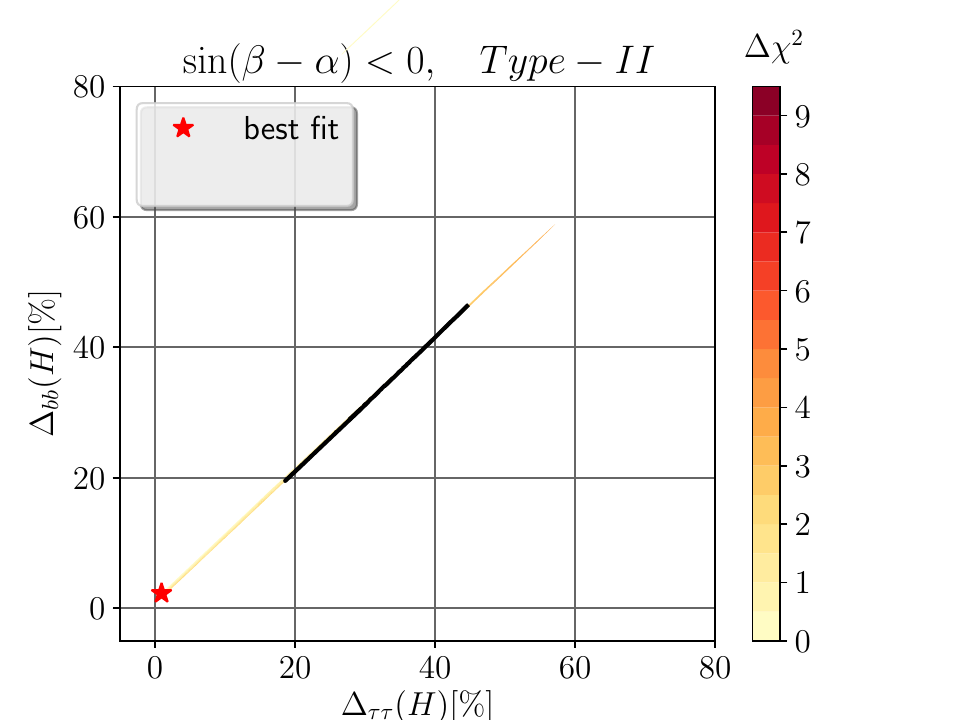}
\includegraphics[width=0.45\textwidth]{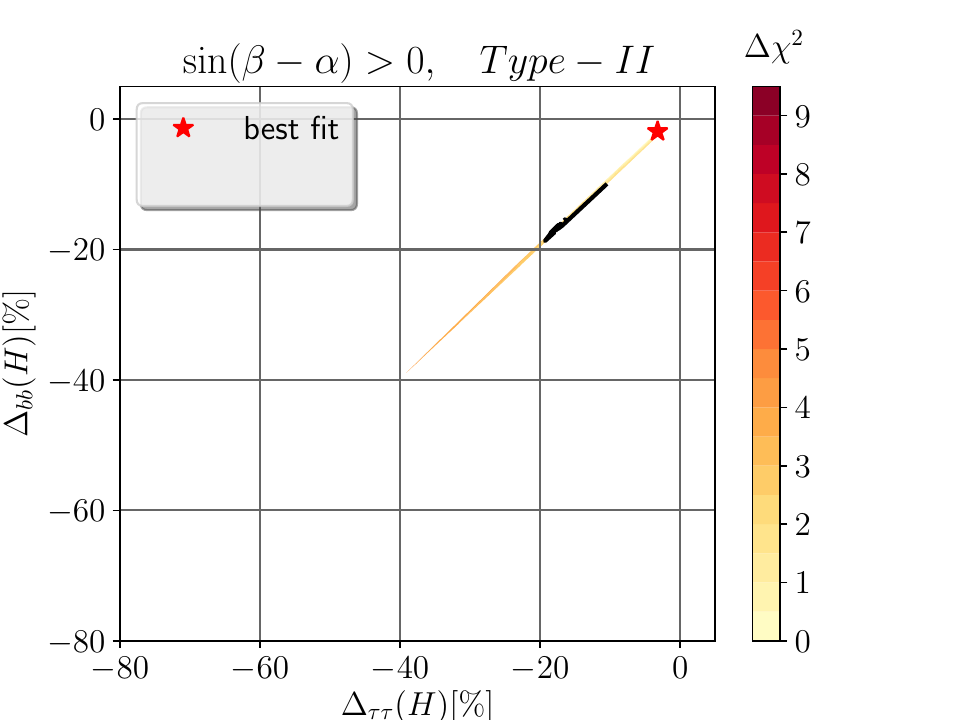}
\end{center}
\caption{Correlation between relative precisions $\Delta_{\tau\tau}(H)$ and
 $\Delta_{bb}(H)$ in 2HDM are shown, where $H^0$ is assumed to be the SM-like Higgs boson. 
 The coding color is the same as in Figure (\ref{fig:best-fit}) and $m_h$ = 95 GeV for the $\pm$D1-H scenarios (upper panel) and the $\pm$D2-H scenarios (lower panel).}
\label{figH:best-fit}
\end{figure}

We use HiggsSignals \cite{Bechtle:2013xfa,Stal:2013hwa,Bechtle:2014ewa} which incorporates Simplified Template Cross section (STXS) framework  \cite{deFlorian:2016spz}  to evaluate the $\chi^2$ of each point in the parameter space. The $\chi^2$ of HiggsSignals we used here includes two parts: 1) a signal strength part $\chi^2_\mu$ and 2) a mass peak part $\sum_{i}\chi^2_{m_i}$. In the signal strength part, when N observables for a given Higgs boson are considered, the $\chi^2_\mu$ (here $\mu$ denotes one of Higgs bosons) equals to $\chi^2_\mu = \sum_{\alpha=1}^N \chi^2_{\mu,\alpha} = (\widehat \mu- \mu )^T C_\mu^{-1} (\widehat \mu - \mu )$, where $(\widehat \mu - \mu )$ should be understood as a N-dimensional vector and $C_\mu$ is the $N\times N$ signal strength covariance matrix. Theoretical observables in the $\chi^2$ of HiggsSignals are defined as signal strengths $\widehat\mu$. When the final states of Higgs decay are specified, a signal strength is defined as $\widehat \mu = \sum_i (\sigma_i^{2HDM} Br^{2HDM})/(\sigma_i^{SM} Br^{SM})$, where $\sigma_i$ denotes the cross section of a production process and $Br$ denotes the decay branching fraction of the Higgs boson. The mass peak part takes into account all neutral Higgs bosons in a model, which is defined as  $\chi^2_{m_i}=\sum_{\alpha=1}^N \chi^2_{m_i,\alpha} = (\widehat m - m_i )^T C_{m_i}^{-1} (\widehat m  - m_i )$, where N neutral Higgs boson is assumed. In our study of 2HDM, the $\chi^2$ mainly takes into account $\chi^2_\mu$ and the $\chi^2_{m_i}$ when a $m_i$ is assigned to be the SM-like Higgs boson. For example, when $H^0$ is assigned to be the SM-like Higgs boson in BP1-H and BP2-H scenarios, the corresponding $\chi^2_\mu$ and $\chi^2_{m_{H^0}}$ are computed.


In Figures (\ref{fig:best-fit}) and (\ref{figH:best-fit}), in order to examine $ \Delta_{ff} (\phi)$ in the 2HDM, we use the mass parameters from Table II and 
  $|\xi^{\phi}_f| = 1$ to fix $\alpha$. Then we allow two parameters $\tan\beta$ and $\lambda_5$ to change. These two parameters are allowed to vary within the range $2 < \tan\beta \le 15 $ and $ -6 \le \lambda_5 \le 1$ while only points which satisfy all theoretical and experimental constraints are accepted. In Figure (\ref{fig:best-fit}), we examine numerically $\Delta_{bb} (h)$ and $\Delta_{\tau\tau} (h)$ for D1-h and D2-h scenarios where $h^0$ is assumed to be the SM-like Higgs boson.  We use D1-h and D2-h to label these two different scenarios which have the same mass spectra of the BP1-h and BP2-h in Table II, respectively. In this practice, it is obvious that the sign of $\cos(\beta - \alpha)$ is allowed to vary, therefore we assume it to be either positive or negative. In Figure  (\ref{figH:best-fit}), we examine $\Delta_{bb} (H)$ and $\Delta_{\tau\tau} (H)$ for D1-H and D2-H scenarios where $H^0$ is assumed to be the SM-like Higgs boson, likewise, D1-H and D2-H scenario share the same mass spectra like BP1-H and BP2-H, respectively. Meanwhile, the sign of $\sin(\beta - \alpha)$ is allowed to be both positive and negative, so we label the scenarios as $\pm$ D1-H and $\pm$ D2-H. Similarly we label other scenarios as $\pm$ D1-h and $\pm$ D2-h. In total, we can have 8 scenarios.


By scanning over the parameter space of 2HDM, we can compute its corresponding $\chi^2$ value. 
Obviously, the $\chi^2$ can be determined for each point in the parameter space and then the difference of $\Delta \chi^2 = \chi^2 - \chi^2_{min}$ can be computed. The total of number of freedom $\nu$ is defined as $ \nu = n_o - n_p$, where $n_o$ is the number of observables taken from experiments measurements which is equal to 78 in this work and $n_p$ is the number of model parameters in 2HDM which is equal to 7. For a general fit, as shown in Figure (\ref{const}) and the results provided in Table \ref{BPdata}, the number of freedom is $78-7=71$. Now, for the specific fits shown in Figures (\ref{fig:best-fit}) and (\ref{figH:best-fit}), due to the constraint of $|\xi^{\phi}_f|=1$ and that we have fixed four mass parameters, there are only two free parameters. Therefore, the degree of freedom is $78-2=76$.

In Figures (\ref{fig:best-fit}) and (\ref{figH:best-fit}) ,  we show points that have satisfied the condition $\Delta \chi^2 \leq  11$ ($<3 \sigma$ for a two-parameter fit).   For all 8 scenarios (labelled as $\pm$Dh-1, $\pm$Dh-2, $\pm$DH-1, and $\pm$DH-2), the main $\Delta \chi^2$ is mainly the contributions of $\chi^2_\mu$ from the signal strength observables since  $\chi^2$ of the mass part is tiny and negligible. Plots are shown up to 99\% C.L. It is remarkable that our results are in good agreement with previous publication \cite{Arhrib:2016snv, Kanemura:2018yai}. There are a few comments :
\begin{itemize}
\item A positive correlation between $\Delta_{\tau\tau}(h)$ and $\Delta_{bb}(h)$ is observed for both $\pm$D1-h and $\pm$D2-h, as demonstrated in Figure  (\ref{fig:best-fit}), where color bar indicates regions close to the minimum $\chi^2$. The best-fit point is indicated by a red star in each of these figures.

It turns out that in the D1-h scenarios, the corrections to $\Delta_{bb} (h)$ and $\Delta_{\tau\tau} (h)$ depend on the sign of $\cos(\beta-\alpha)$, i.e. when $\cos(\beta - \alpha) > 0$ ($\cos(\beta - \alpha) < 0$) the corrections are positive (negative). It can deviate from the prediction of the SM by $+15\%$($-6\%$) for the best fit point. In contrast, it is opposite for the BP2-h scenario and allowed to deviate from the prediction of the SM by less than $-6\%$($+5\%$).

\item An interesting observation is that due to the positive correlation between $\Delta_{bb} (h)$ and $\Delta_{\tau\tau} (h)$, it is possible to determine the sign of $\cos(\beta-\alpha)$ by using a precise measurements of the branching fractions of $h \to b\bar{b}$ and $h \to \tau\tau$. As pointed out in Refs.~\cite{Dawson:2013bba,Zeppenfeld:2000td,Gianotti:2000tz,Arhrib:2016snv}, if the couplings of Higgs to fermions can be determined up to $8\%$, it becomes possible to discriminate the types of Yukawa interactions. 

\item At the CEPC or the ILC \cite{CEPCStudyGroup:2018ghi, Moortgat-Picka:2015yla, Bambade:2019fyw}, the precision of Higgs coupling measurements can reach typically ${\cal O}$(1)$\%$, and there is no doubt on their capability to discriminate the types of Yukawa interactions, like 4 scenarios introduced here.

\end{itemize}

The deviations of H decaying into $b\bar{b}$ and $\tau\tau$ are presented in Figure (\ref{figH:best-fit}), where the correlations of $\Delta_{bb} (H)$ and $\Delta_{\tau\tau} (H)$ for $\pm$D1-H and $\pm$D2-H scenarios are demonstrated. One remarkable difference from Figure (\ref{fig:best-fit}) is that the allowed deviation of $\Delta_{bb} (H)$ and $\Delta_{\tau\tau} (H)$ from those of the SM is smaller than $2.5\%$ ($-3\%$) for the best fit point. Such a deviation is even smaller (less than $1\%$) as shown by the BP2-H in the D2-H scenarios. Such deviations are quite distinctive due to Yukawa structure in these four scenarios. In other words, when a deviation is observed, an upper limit can be set on the mixing angles $\cos(\beta - \alpha)$ or $\sin(\beta-\alpha)$.

\section{Discussions and Conclusion}

In this work, we have proposed 4 benchmark scenarios of the 2HDM after taking into account the current Higgs data from the LHC and evaluated the radiative corrections to the process $e^+ e^- \to Z \phi$ in the SM and these 4 benchmark scenarios up to one-loop level.

In the Monte Carlo simulation results \cite{Mo:2015mza} obtained using Whizard, it is found that the ISR  can decrease the total LO cross-section of $e^+ e^- \to Z h$ in the SM by more than $10\%$ when the contribution of high order logarithms are resummed. In contrast, the real emission in this work can increase the total LO cross-section by only a factor $+0.5\%$. This simply manifests the difference between the resummed and the fixed order results. Meanwhile, we find that the 1-loop weak correction of the SM can reduce the total NLO cross section by $9.15\%$, as demonstrated in Table \ref{nlo_SM}.  

As an estimation, we can express the total cross section with ISR as 
\begin{equation}
\sigma^{LO}(s) ={\bar \sigma^{W,LO}} (s) R(1 - \frac{\Delta E}{E}, s)\,,
\end{equation}
which had been proposed by the reference \cite{Kuraev:1985hb} when $\frac{\Delta E}{E}<< 1$ ($E$ denotes the energy of electron/positron, and $\Delta E$ is the energy of soft photon), where $R(1 - \frac{\Delta E}{E}, s)$ denotes the convolution of incoming fluxes of electron and positron, i.e. ISR, and ${\bar \sigma^{LO}} (s)$ denotes the cross section without weak corrections. It should be pointed out that ${\bar \sigma^{W,LO}}$ has included high order QED corrections but no weak corrections. The numerical results given in \cite{Mo:2015mza} demonstrated that $R(1 - \frac{\Delta E}{E}, s)$ is roughly equal to $0.90$. 

When the one-loop weak correction is taken into account, the total cross section at NLO with ISR can be expressed as 
\begin{equation}
\sigma^{NLO}(s) = {\bar \sigma^{W,NLO}} (s) R(1 - \frac{\Delta E}{E}, s)
\end{equation}
where the $R(1 - \frac{\Delta E}{E}, s)$ (i.e. ISR) is the same as in the LO case. Our results demonstrate that weak correction can reduce ${\bar \sigma^{W,LO}}$ by a factor $9.15\%$ with $\mu=m_Z$. We can use the NLO cross section ${\bar \sigma^{W,NLO}}=228$ fb with $\mu=m_Z$ to estimate the total cross section with both ISR and weak corrections, which yields $\sigma^{NLO}(s) = 228  \times 0.9 = 205.2$ fb, which is $18.6\%$ smaller than the tree level cross section in the $\alpha_{em}(m_Z)$ scheme given in Table \ref{nlo_SM}.

Here we add one more comment on the ISR. As it is well-known that due to the soft-collinear divergences of QED, the large logarithmic terms in the form of $\alpha_{em}^n \log^n\frac{s}{m_e^2} \log^m\frac{m_e}{\lambda}$ (here $\lambda$ denotes the mass of photon or IR cutoff) from the ISR should be resummed. The effects of ISR at electron-positron colliders can affect cross-sections of physics processes significantly \cite{Berends:1984dw,Jadach:1999pp}.
In Whizard \cite{Kilian:2007gr}, the effect of ISR to all orders \cite{Gribov:1972rt,Kuraev:1985hb,Skrzypek:1990qs} 
 has been implemented in the structure functions of incoming electron and positron. The ISR has been implemented in other Monte Carlo codes \cite{Barberio:1993qi,Jadach:1999pp,Jadach:1999vf,Skrzypek:1995wd,Jadach:2001mp} and applied to LEP I \cite{Barate:1999ce} and LEP II \cite{Schael:2013ita} experimental analysis. Therefore, it is expected that the results for the SM from Whizard with ISR is more accurate in capturing large logs. While for the study of the 2HDM, currently, ISR effect has not implemented in Whizard and therefore there is no alternative code for us to check it.
 
We compute the decays $\phi \to b\bar{b}$ and $\phi\to \tau^+ \tau^-$ with $\phi = h^0$ ( $H^0$ ) in Dh-1 and Dh-2 (DH-1 and DH-2) by including the EW corrections. We have shown that in Dh-2 and DH-2 scenarios, the electroweak radiative corrections in these two decay processes are rather small due to the fact that the heavy states $A^0$, $H^0$ and $H^\pm$ have masses of order of 600 GeV while they could be sizeable for D1-h and D1-H, as shown in Figure  (\ref{fig:best-fit}) and Figure  (\ref{figH:best-fit}). Considering the recent progress in the gauge independent renormalization schemes \cite{Krause:2016oke, Denner:2016etu,Altenkamp:2017ldc}, it will be interesting to evaluate the differences caused by different renormalization schemes.

Our results demonstrate that $e^+e^-$ colliders (especially the Higgs factories with $\sqrt{s}=250$ GeV) can offer us the potential to distinguish various 2HDM models by looking at quantum effects in Higgs observables. Except for performing precision measurements of the SM-like Higgs boson, linear colliders also have the potential to discover new physics. For example, it can directly produce the light charged Higgs boson pair in BP1-h and BP1-H scenarios via $e^+ e^- \to H^+ H^-$ processes \cite{Komamiya:1988rs,Arhrib:1998gr}. The energy scan of $e^+e^-$ colliders can also help to detect the mass spectra of Higgs bosons and even triple couplings of Higgs bosons, as shown in Figure  (\ref{fig1:weakthdm3}) and Figure  (\ref{fig2:weakthdm4}). If nature chooses 2HDM as the new physics beyond the SM at TeV region, and it would be hopeful to probe parameters of the Higgs potential sector by using the data of productions and decays. The results of the proposed benchmark points show that precision measurement on the cross-sections at different collision energy is helpful to explore the mass spectra and Higgs couplings of the 2HDM.

\section*{Acknowledgments}

We thank Abdesslam Arhrib and David Lopez-Val for helpful discussion at the early stage of this work. This work is supported by the Moroccan Ministry of Higher
Education and Scientific Research MESRSFC and CNRST: Project PPR/2015/6.
RB was supported in part by the Chinese Academy of Sciences (CAS) President, International Fellowship Initiative (PIFI) program (Grant No. 2017VMB0021). Q.S.Yan is supported by the Natural Science Foundation of China under grant No. 11475180 and No. 11875260.
B. Gong is supported by the Natural Science Foundation of China No. 11475183.


\appendix

\section{The IR beheavior of $e^+ e^- \to Z h$ in the SM}
\label{checkingpoints}
The dependence of the corrections at NLO, $\sigma^{1}$, on $\Delta E$ and $\Delta\theta$ are shown in Tables \ref{check:e} and \ref{check:theta}. In Table \ref{check:e}, the dependence is seen in a wide range and we choose $\delta_s=10^{-3}$ as our default choice. In Table \ref{check:theta}, the result becomes cut dependent when $\Delta\theta$ is smaller than $10^{-4}$. It is because the approximation in Eq.~(\ref{eqn:HC}) demands $\Delta\theta\gg m_e/\sqrt{s}\sim 2\times10^{-6}$. Thus we choose $\Delta\theta=10^{-3}$ in our setting.

\begin{table}[http]
\begin{center}
\begin{tabular}{|c|c|c|c|c|c|c|}
\hline\hline
$\delta_s=2\Delta E/\sqrt{s}$&$\sigma_{S+V}$&$\sigma_{H\overline{C}}$&$\sigma_{HC+CT}^{*}$&$\sigma_{SC}$ & $\sigma^{1}$\\
\hline
$10^{-1}$&-0.7127(0)&0.1240(0)& -0.1209(0)&0.4794(0)&-0.2302(0)\\
$10^{-2}$&-1.4347(0)&0.5445(0)&-0.5306(1)&1.1903(0)&-0.2305(1)\\
$10^{-3}$&-2.1567(0)&0.9788(1)&-0.9540(2)&1.9012(0)&-0.2307(2)\\
$10^{-4}$&-2.8787(0)&1.4142(1)&-1.3784(2)&2.6121(0)&-0.2308(2)\\
$10^{-5}$&-3.6006(0)&1.8497(2)&-1.8027(4)&3.3230(0)&-0.2306(4)\\
$10^{-6}$&-4.3227(0)&2.2853(2)&-2.2271(5)&4.0339(0)&-0.2306(5)\\
$10^{-7}$&-5.0446(0)&2.7208(2)&-2.6516(6)&4.7448(0)&-0.2306(6)\\
$10^{-8}$&-5.7666(0)&3.1564(3)&-3.0762(7)&5.4558(0)&-0.2306(8)\\
 \hline
\end{tabular}
\caption{Check for $\Delta E$ independence at $\sqrt{s} = 250$ GeV (in unit of $10^{-1}$ pb).}
\label{check:e}
\end{center}
\end{table}

\begin{table}[http]
\begin{center}
\begin{tabular}{|c|c|c|c|c|c|}
\hline\hline
$\Delta\theta$&$\sigma_{V+S}$&$\sigma_{H\overline{C}}$&$\sigma_{HC+CT}^{*}$&$\sigma_{SC}$& $\sigma^{1}$ \\
\hline
$10^{-1}$&-2.1567(0)&0.3856(1)&-0.3608(1)&1.9012(0)&-0.2307(1)\\
$10^{-2}$&-2.1567(0)&0.6822(1)&-0.6574(1)&1.9012(0)&-0.2307(1)\\
$10^{-3}$&-2.1567(0)&0.9788(1)&-0.9540(2)&1.9012(0)&-0.2307(2)\\
$10^{-4}$&-2.1567(0)&1.2751(1)&-1.2506(3)&1.9012(0)&-0.2310(3)\\
$10^{-5}$&-2.1567(0)&1.5527(1)&-1.5471(3)&1.9012(0)&-0.2499(3)\\
$10^{-6}$&-2.1567(0)&1.6227(2)&-1.8437(4)&1.9012(0)&-0.4765(4)\\
 \hline
\end{tabular}
\caption{Check for $\Delta\theta$ independence at $\sqrt{s} = 250$ GeV (in unit of $10^{-1}$ pb).}
\label{check:theta}
\end{center}
\end{table}

One-loop radiation corrections include collinear singularities which can be infinite when $m_e$ goes to zero, and they become terms proportional to $\log(m_e)$ in this limit.
Some of these terms are cancelled when virtual and real corrections are summed up,
some of them can be absorbed into the redefinition of running coupling constant as mentioned above,
but the collinear singularities can not be removed completely, which demands a careful manipulation.

To deal with these remaining collinear singularities, we used following fixed order electron structure function which can be derived from Eq.~(11) of Ref.~\cite{Kuraev:1985hb}:
\begin{equation}
\label{eqn:ff}
f_{ee}(x,s)=\delta(1-x)+\dfrac{\alpha_{em}}{2\pi}\log\dfrac{s}{4m_e^2}P_{ee}^+(x,0)
\end{equation}
with
\begin{equation}
P_{ee}^+(z,0)=\dfrac{1+z^2}{(1-z)_+}+\dfrac{3}{2}\delta(1-z),
\end{equation}
being the regularized Altarelli-Parisi splitting function.
One-loop structure function is used here to ensure the cancellation of collinear singularities, instead of using most-commonly-used resummed ones. The cancellation is shown in Table \ref{check:me}.
We vary the mass of electron with a factor of $k$ from $2^{-4}$ to $2^8$ and find the result unchanged. Also, we can see that singular terms only appear in $\sigma_{V+S}$ and $\sigma_{SC}$ parts.
\begin{table}[http]
\begin{center}
\begin{tabular}{|c|c|c|c|c|c|c|}
\hline\hline
$k$&$\sigma_{V+S}$&$\sigma_{H\overline{C}}$&$\sigma_{HC+CT}^{*}$&$\sigma_{SC}$&$\sigma^{1}$\\
\hline
$2^{-4}$ &-2.5815(0)&0.9788(1)&-0.9540(2)&2.3260(0)&-0.2307(2)\\
$2^{-3}$ &-2.4753(0)&0.9788(1)&-0.9540(2)&2.2198(0)&-0.2307(2)\\
$2^{-2}$ &-2.3691(0)&0.9788(1)&-0.9540(2)&2.1136(0)&-0.2307(2)\\
$2^{-1}$ &-2.2629(0)&0.9788(1)&-0.9540(2)&2.0074(0)&-0.2307(2)\\
$2^0$   &-2.1567(0)&0.9788(1)&-0.9540(2)&1.9012(0)&-0.2307(2)\\
$2^1$   &-2.0505(0)&0.9788(1)&-0.9540(2)&1.7950(0)&-0.2307(2)\\
$2^2$   &-1.9443(0)&0.9788(1)&-0.9540(2)&1.6888(0)&-0.2307(2)\\
$2^3$   &-1.8381(0)&0.9788(1)&-0.9540(2)&1.5826(0)&-0.2307(2)\\
$2^4$   &-1.7318(0)&0.9788(1)&-0.9540(2)&1.4763(0)&-0.2307(2)\\
$2^5$   &-1.6256(0)&0.9788(1)&-0.9540(2)&1.3701(0)&-0.2307(2)\\
$2^6$   &-1.5194(0)&0.9788(1)&-0.9540(2)&1.2639(0)&-0.2307(2)\\
$2^7$   &-1.4132(0)&0.9788(1)&-0.9540(2)&1.1577(0)&-0.2307(2)\\
$2^8$   &-1.3070(0)&0.9788(1)&-0.9540(2)&1.0515(0)&-0.2307(2)\\
 \hline
\end{tabular}
\caption{Check for $m_e$ independence at $\sqrt{s} = 250$ GeV (in unit of $10^{-1}$ pb). }
\label{check:me}
\end{center}
\end{table}

\section{Triple Higgs couplings in 2HDM}\label{trihiggsc}
The Higgs potential of 2HDM determines the self-couplings among Higgs bosons. Among them, the triple Higgs couplings (THC)
can be parameterized as a function of the 2HDM parameters
$m_{h^0}$, $m_{H^0}$, $m_{A^0}$, $m_{H^\pm}$, $\tan\beta$, $\alpha$
and $\lambda_5$.
At tree level, these couplings are independent of the Yukawa types and they are given as follows

\begin{eqnarray}
&\lambda_{h^0h^0h^0}^{2HDM} = \dfrac{1}{v}\left\{-\frac{3}{s^2_{2\be}}\bigg[(2c_{\al+\be} + s_{2\al}s_{\be-\al})s_{2\be} m^2_{h^0} - 4c^2_{\be-\al} c_{\be + \al} m^2_{12}\bigg] \right\}\label{lll} \\
&\lambda_{H^0h^0h^0}^{2HDM} = \dfrac{1}{v}\left\{-\frac{c_{\be-\al}}{s^2_{2\be}}\bigg[
  (2 m^2_{h^0} + m^2_{H^0}) s_{2\al} s_{2\be} -2 (3 s_{2\al}-s_{2\be})
  m^2_{12}\bigg] \right\}  \label{Hll} \\
&\lambda_{h^0H^0H^0}^{2HDM} = \dfrac{1}{v}\left\{\frac{s_{\be-\al}}{s^2_{2\be}}\bigg[
  (m^2_{h^0} + 2 m^2_{H^0}) s_{2\al} s_{2\be} - 2 (3 s_{2\al}+s_{2\be})
  m^2_{12}\bigg] \right\} \label{hhl}\\
& \lambda_{ h^0H^\pm H^\mp}^{2HDM} = \dfrac{1}{v}\left\{\frac{1}{s^2_{2\be}}\bigg[
  (m^2_{h^0} - 2 m^2_{H^\pm})s_{\be-\al}s^2_{2\be} -
2c_{\be + \al}(m^2_{h^0} s_{2\be}-
 2 m^2_{12})\bigg] \right\}  \label{hhp}\\
& \lambda_{h^0A^0A^0}^{2HDM}  = \dfrac{1}{v}\left\{\frac{1}{s^2_{2\be}}\bigg[
  (m^2_{h^0} -2 m^2_{A^0} )s_{\be-\al}s^2_{2\be} - 2c_{\be + \al}(m^2_{h^0} s_{2\be}-
  2 m^2_{12})\bigg] \right\}\label{lambda5}
\\& \lambda_{H^0H^0H^0}^{2HDM}=\dfrac{1}{v}\left\{-\frac{3}{s^2_{2\be}} \bigg[(2 s_{\al+\be}-s_{2 \alpha} c_{\beta-\alpha})s_{2\be}m_{H^{0}}^2-4s^2_{\beta-\alpha} s_{\alpha+\beta}m_{12}^2\bigg] \right\}\label{HHH}
\\ &\lambda_{H^0A^0A^0}^{2HDM}=\dfrac{1}{v}\left\{-\frac{1}{s^2_{2\be}}\bigg[s_{\alpha+\beta} (2 m_{H^0}^2 s_{2 \beta}-4 m_{12}^2)-(m_{H^0}^2-2 m_{A^0}^2)s^2_{2\beta} c_{\beta-\alpha}\bigg] \right\}\label{HAA}
\\ &\lambda_{H^{\pm}H^{\mp}H^0}^{2HDM}=\dfrac{1}{v}\left\{-\frac{1}{s^2_{2\be}}\bigg[s_{\alpha+\beta} (2 m_{H^{0}}^2 s_{2\beta}-4 m_{12}^2)-(m_{H^0}^2-2 m_{H^\pm}^2)s^2_{2 \beta}c_{\beta-\alpha}\bigg]\right\} \label{hpH}
\end{eqnarray}
where $v$ is the VEV and $m_{12}^2$ can be derived from $m_A$, $\beta$, $\lambda_5$ and $v$, according to the relation given in Eq.~(\ref{l5m122}).
We have used the notation $s_\theta$ and $c_\theta$ as short-hand notations for
$\sin(\theta)$ and $\cos(\theta)$, respectively.
 The mixing angle $\beta$ is defined by via $\tan\beta=v_2/v_1$.

 As mentioned in the second paragraph of section \ref{sec:results}, two different physical parameterizations have been used in this paper, which yields different values of $v$ and hence different THCs. Besides the overall factor $1/v$, the remaining terms of Eqs.~(\ref{lll}-\ref{hpH}) still depend on $v$,  due to the fact that $m_{12}^2$ is obtained via Eq.~(\ref{l5m122}) and terms proportional to $\lambda_5 v^2$ will appear.
 In order to further investigate the dependence, we rewrite the THCs as:
 \begin{equation}
\lambda_i \equiv \dfrac{(246.220~\mathrm{GeV})^2} {v} \hat{\lambda}_i , \quad i=h^0h^0h^0, H^0h^0h^0, \dots
\label{eqn:hl3h}
\end{equation}
where the dimensionless couplings $\hat{\lambda}_i$ are just the terms in the curly brackets of Eqs.~(\ref{lll}-\ref{hpH}) divided by (246.220 GeV)$^2$. 
 
 \begin{table}[http]
\begin{center}
 \begin{tabular}{|c|c| r| r | r | r | r |r|r|r|r|}
        \hline\hline
       $v$(GeV)&BPs& $\hat{\lambda}_{h^0h^0h^0}$ & $\hat{\lambda}_{H^0h^0h^0}$ & $\hat{\lambda}_{h^0H^0H^0}$ & $\hat{\lambda}_{h^0H^\pm H^\pm}$ &  $\hat{\lambda}_{h^0A^0A^0}$ &$\hat{\lambda}_{H^0H^\pm H^\pm}$ & $\hat{\lambda}_{H^0A^0A^0}$&$\hat{\lambda}_{H^0H^0H^0}$&$m_{12}^2$(GeV$^2$)\\
        \hline
       \multirow{4}*{246.220} 
       &BP1-h & -0.763 & -0.048& -0.810 &-0.375 & 0.353  &-0.030 &0.028&-0.195&3126.00\\ \cline{2-11}
       &BP2-h & -0.773 &-0.049 & -3.257 & -3.199 & -0.285 &-2.305&-2.292&-6.917&104480.72\\\cline{2-11}
       &BP1-H & 0.025 & -0.240 & -0.002 & 0.048 & 0.047 &-0.895&-0.884&-0.774&2839.78\\\cline{2-11}
       &BP2-H & -0.036 & -0.235 & -0.001 & 0.335 & 0.335 &-11.807&-11.807&-0.773&3567.16\\
       \hline
        \multirow{4}*{243.137} 
         &BP1-h & -0.763 & -0.047& -0.789 &-0.388 & 0.340  &-0.238 &-0.179&-0.814&3064.92\\\cline{2-11}
         &BP2-h & -0.773 &-0.049 & -3.257 & -3.199 & -0.285 &-2.305&-2.292&-6.917&104480.72\\\cline{2-11}
         &BP1-H & -0.037 & -0.228 & -0.0003 & 0.027 & 0.026 &-0.880&-0.869&-0.774&2992.36\\ \cline{2-11}
         &BP2-H & -0.793 & 0.030 & 0.016 & 0.083 & 0.083 &-11.527&-11.527&-0.773&6981.21\\
        \hline\hline
      \end{tabular} 
      \caption{Dimensionless couplings $\hat{\lambda}_i$ with two typical values of $v$, the inputs of each of benchmark points are given in Table \ref{BPdata}. $\hat{\lambda}_i$ are defined from the THCs via Eq.~(\ref{eqn:hl3h}). The values of $v$ are taken as explained in the second paragraph of Section \ref{sec:results}.}   \label{thc-num}
  \end{center}
    \end{table}

\begin{figure}[http]
\centering
\subfigure[ BP1-h]{\includegraphics[width=0.4\textwidth]{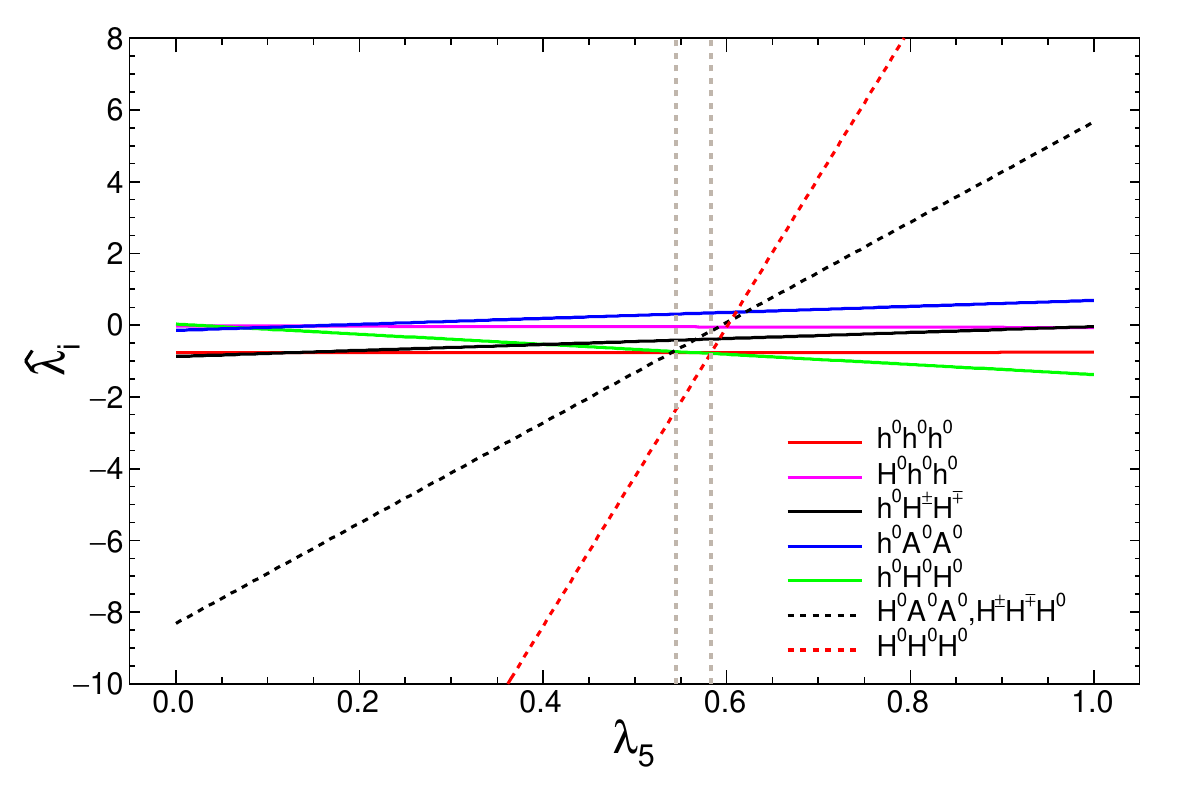}}
\subfigure[ BP2-h]{\includegraphics[width=0.4\textwidth]{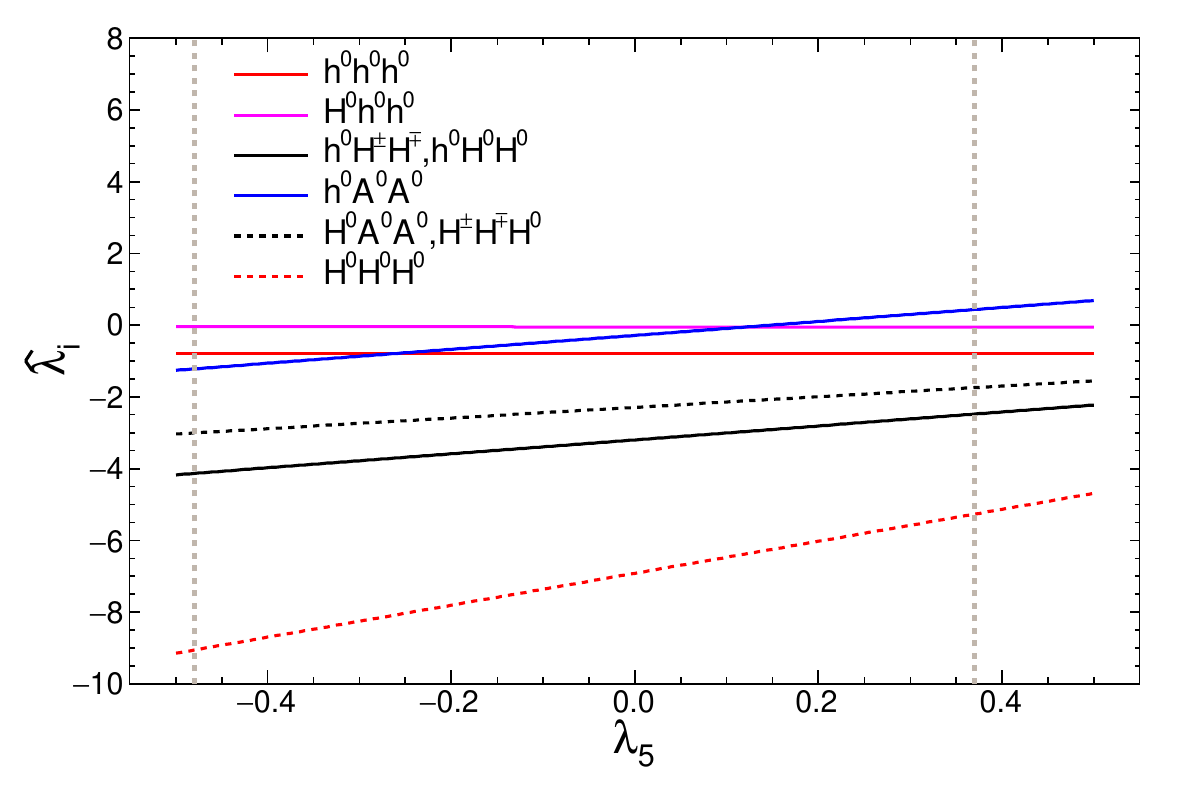}}
\subfigure[ BP1-H]{\includegraphics[width=0.4\textwidth]{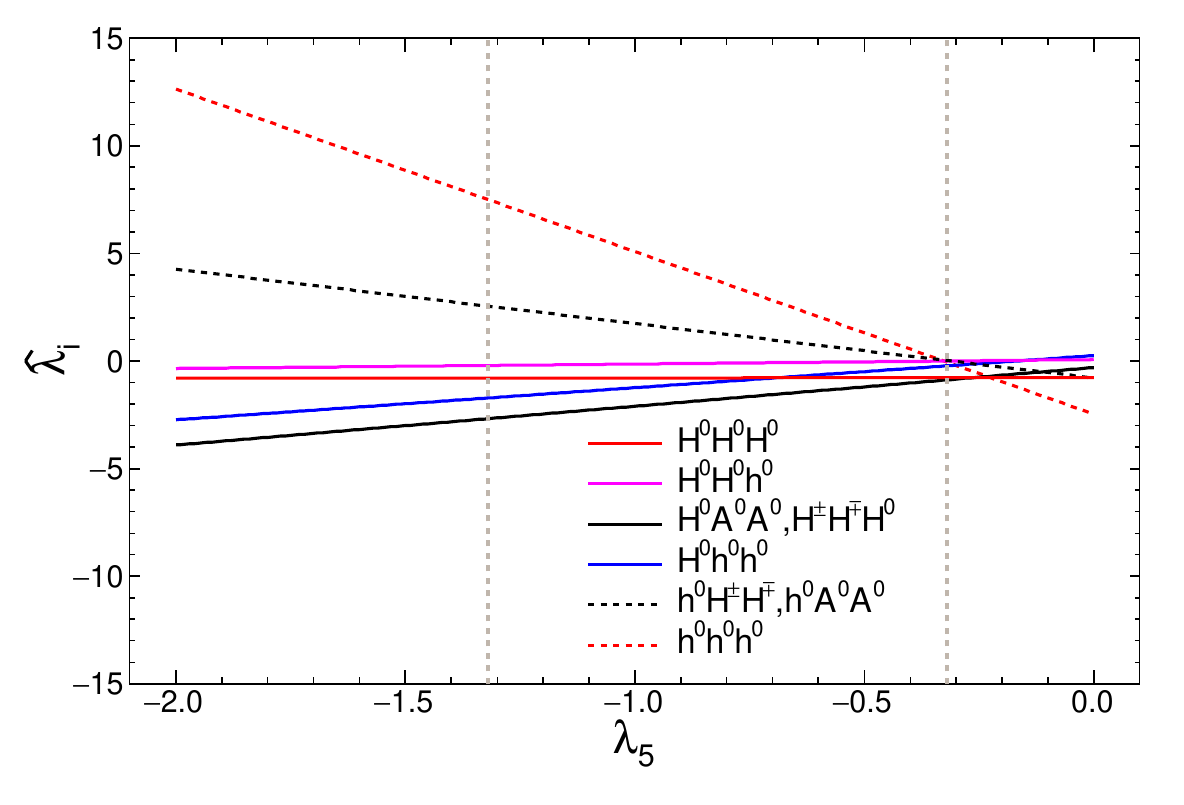}}
\subfigure[ BP2-H]{\includegraphics[width=0.4\textwidth]{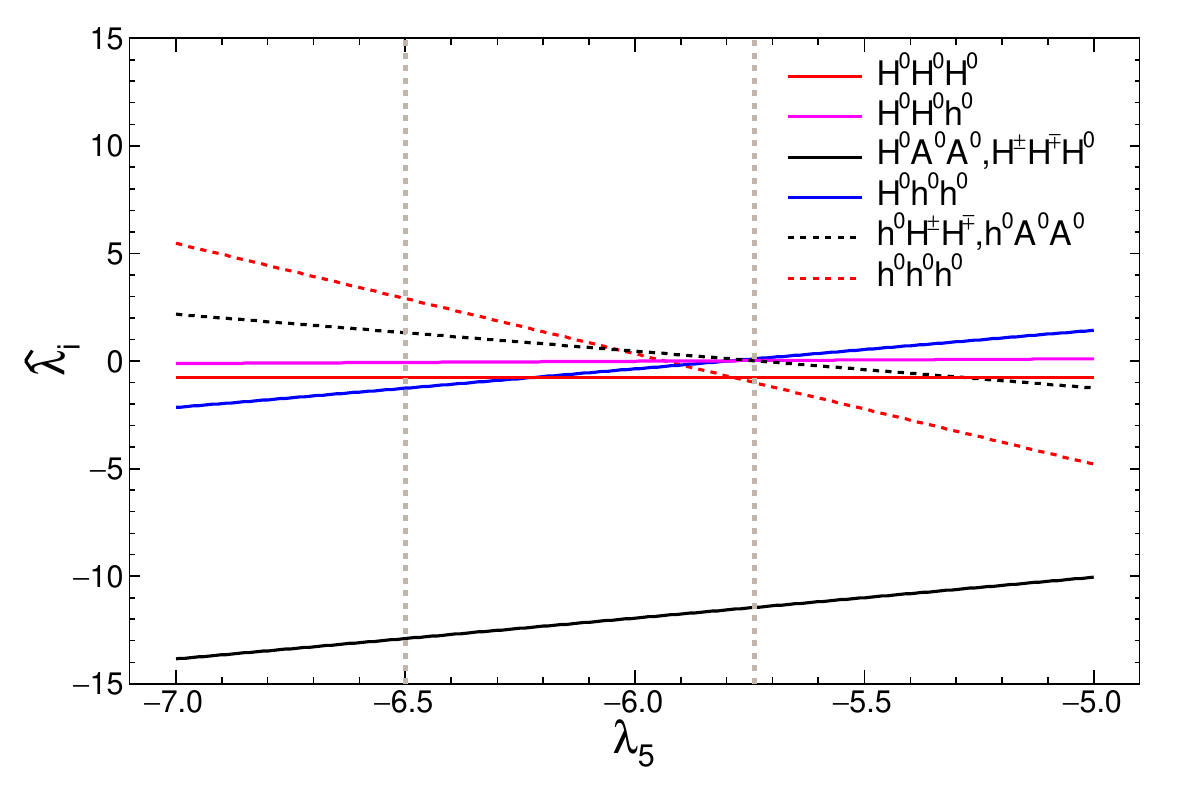}}
\caption{The dependence of $\hat{\lambda}_i$ upon $\lambda_5$ for each of benchmark points is demonstrated. In each plot, the range between two vertical dashed lines are the allowed $\lambda_5$ by theoretical and experimental constraints and bounds. $v$ is taken as 243.137 GeV.}
\label{figthc:thcs}
\end{figure}

In Table \ref{thc-num}, all the $\hat{\lambda}_i$ in the two cases are tabulated, as well as corresponding values of $m_{12}^2$. It can be seen from the table that in the case of BP2-h (BP2-H), some of the THCs are quite large compared with $h^0 h^0 h^0$ ($H^0 H^0 H^0$). Although these THCs are sizeable, we have checked that they still  respect the conditions of perturbativity and the unitarity of scattering amplitudes of Higgs bosons in the model, as listed on page 3. 

On the other hand, although the deviation between the two VEVs are smaller than 1.5\%, it is observed that the values of $\hat{\lambda}_i$ can be significantly different, such as $\hat{\lambda}_{H^0H^\pm H^\pm}$, $\hat{\lambda}_{H^0A^0A^0}$ and $\hat{\lambda}_{H^0H^0H^0}$ in the case of BP1-h, and $\hat{\lambda}_{h^0h^0h^0}$, $\hat{\lambda}_{H^0h^0h^0}$, $\hat{\lambda}_{h^0H^\pm H^\pm}$ and $\hat{\lambda}_{h^0A^0A^0}$ in the case of BP2-H, which indicates the corresponding THCs are very sensitive to the value of $v$. As $v$ only appears in $\hat{\lambda}_i$ in terms of $\lambda_5 v^2$, this also means those THCs are very sensitive to the value of $\lambda_5$ (a deviation smaller than 3\% will change the corresponding $\hat{\lambda}_i$ drastically). Usually the effects caused by the different values of $v$ are regarded as higher order corrections. 
But from Table \ref{thc-num} we can see that in some cases this will lead to totally different theoretical predictions, e.g. the decay width of $H^0\rightarrow A^0A^0$ in the case of BP1-h. Fortunately in this work, this difference in $\hat{\lambda}_i$ does not change our results very much (see also the discussion below).

In Figure  (\ref{figthc:thcs}), we show the dependence of $\hat{\lambda}_i$ upon the parameter $\lambda_5$ for each of benchmark mass spectra and  fixed parameters $\alpha$ and $\tan\beta$ given in Table \ref{BPdata}, where $v=243.137$ GeV is used.
In the benchmark points BP1-h(H) and BP2-h(H) where $h^0$($H^0$) is assumed to be the SM-like Higgs boson, the $\hat{\lambda}_i$ related to $h^0$($H^0$) are depicted as solid lines  while the others are depicted by dashed lines as they can also contribute via the renormalization of wave functions or mixing angles.
We have merged several couples of lines in the figure since they are identical or very close to each other, which can be easily learned from Eqs.~(\ref{lll}-\ref{hpH}) and the mass spectra in Table~\ref{BPdata}. 

The range of $\lambda_5$ between two vertical dashed lines denotes the allowed region for each benchmark point. 
It is observed from the figure that the gradients of some lines are quite large, thus the corresponding $\hat{\lambda}_i$ change drastically in the allowed range of $\lambda_5$, such as $\hat{\lambda}_{h^0h^0h^0}$ in the case of BP1-H which changes from $7.5$ to $-0.05$. As we discussed before, the dependence of $\hat{\lambda}_i$ on $v$ can be converted into the dependence on $\lambda_5$. This explains the discrepancy of some $\hat{\lambda}_i$ caused by the different values of $v$ in Table \ref{thc-num}. Also, it can be estimated that the effect caused by different values of $v$ (after removing the overall factor) is much smaller than the variation of $\lambda_5$ in the allowed range, since the difference between the values of $v$ can be taken as a deviation smaller than $3\%$ on $\lambda_5$.

 \begin{table}[http]
\begin{center}
 \begin{tabular}{|c|c|c|c|c|c|}
        \hline\hline
              $v$(GeV)&BPs  &~~~~~~~~BP1-h~~~~~~~~&~~~~~~~~BP2-h~~~~~~~~&~~~~~~~~BP1-H~~~~~~~~&~~~~~~~~BP2-H~~~~~~~~\\
        \hline
       \multirow{3}*{246.220} 
       &$\lambda_5$ & [0.545,0.583] & [-0.482,0.377]& [-1.330,-0.294] &[-6.500,-5.737] \\ \cline{2-6}
       &$\lambda_5v^2$(GeV$^2$)& [33040, 35368] &[-29197, 22874]& [-80612, -17848] &[-394063, -347807]\\\cline{2-6}
       &$m_{12}^2$(GeV$^2$)&[2970, 3132] &[92844, 113597]&[-16351, 3365]&[-12986, 4649] \\\cline{2-6}
               \hline    
       \multirow{3}*{243.137} 
       &$\lambda_5$ & [0.560, 0.598] &[-0.441, 0.387]&[-1.337, -0.302]&[-6.601, -5.884]       \\ \cline{2-6}
       &$\lambda_5v^2$(GeV$^2$)&[33099, 35367]&[-26094, 22872]&[-79032, -17847]&[-390217, -347807]\\\cline{2-6}
       &$m_{12}^2$(GeV$^2$)&[2974, 3132]&[94081, 113597]&[-15855, 3366]&[-11520, 4649] \\\cline{2-6}
        \hline\hline
      \end{tabular} 
      \caption{\label{tab:l5range}Allowed ranges of $\lambda_5$, $\lambda_5v^2$ and $m_{12}^2$ with two typical values of $v$. The constraints from vacuum stability, unitarity and perturbativity are used to determine the ranges. The values of $v$ are taken as explained in the second paragraph of Section \ref{sec:results}.}   
  \end{center}
    \end{table}

We have discussed the dependence of THCs on $v$. In fact, the allowed ranges of $\lambda_5$ also depend on it. 
In Table \ref{tab:l5range}, we show the allowed ranges of $\lambda_5$ for all the BPs.
As mentioned in Section \ref{sec:results}, we determine the allowed ranges of $\lambda_5$ by fixing other parameters and using  the constraints from vacuum stability, unitarity and perturbativity. In this procedure we have taken $v=246.220$ GeV, same as the one used in the scan of the 2HDM parameter space.
However in the calculation of production processes, $v=243.137$ GeV is used.

It can be observed from Table \ref{tab:l5range} that both the upper and lower bounds of $\lambda_5$ will change when a different $v$ is used. 
The largest deviation is found in the lower bound of BP2-h, which is about 10\%. This can not be simply explained by an overall factor related to $v$ only, and needs further investigation.

The constraints used to determine the ranges of $\lambda_5$ can be expressed as several dimensionless inequalities involving $v$ and the 2HDM parameters: $m_{h^0}$, $m_{H^0}$, $m_{A^0}$, $m_{H^\pm}$, $\tan\beta$, $\alpha$ and $\lambda_5$. Besides an overall factor $1/v^2$, the only dependence on $v$ in these inequalities is the term $\lambda_5 v^2$, similar to the case of THCs. This means we can remove part of the effects from $v$ by looking into $\lambda_5 v^2$ or $m_{12}^2$. 
In Table \ref{tab:l5range}, the ranges of $\lambda_5 v^2$ and $m_{12}^2$ are also presented. It is found that the upper bounds of $\lambda_5 v^2$ and $m_{12}^2$ are almost same even different values of $v$ are used, while the lower bounds are quite different. We have checked the constraints related to each bound and found that all the upper bounds are determined by the constraints from vacuum stability while all the lower bounds are determined by the constraints from unitarity. The constraints from vacuum stability are homogeneous so that they are not affected by the choice of $v$. On the other hand the constraints from unitarity are inhomogeneous, and some of them are even nonlinear. 
This leads to more complicated $v$ dependence of $\lambda_5 v^2$ and $m_{12}^2$. 
From the table we can see that the effects of different $v$ on the lower bounds can be negligible such as $m_{12}^2$ in BP1-h, but they can also be large such as $m_{12}^2$ in BP2-H. 

From above analysis we can conclude that deviations of the upper bounds of $\lambda_5$ can be easily understood as by multiplying the ratio of two different $v^2$, while the lower bounds are more complicated. 
What's more, together with the situation of THCs, it hints that it is better to choose independent physical parameters with same mass dimension.

\section{On-shell pinched tadpole scheme in the 2HDM\label{appendixc}}
In this appendix we present more details about the on-shell PTS~\cite{Krause:2016oke,Krause:2016xku}, which is used in the calculation of the production $e^+e^-\rightarrow Z\phi$. For the study on $\phi \to f\bar{f} $ decay, as it is a work following Ref.~\cite{Arhrib:2004ak}, the same renormalization scheme as in Ref.~\cite{Arhrib:2004ak} is used and will not be discussed here.

\subsection{Definition of renormalization constants}
In the following we introduce renormalized quantities and renormalization constants in the on-shell PTS, where an index 0 is used to label bare quantities. 
\begin{itemize}
\item Gauge and fermion sectors: things are similar as in the SM for these two sectors. 
Renormalization of gauge boson and fermion masses are given by
\begin{eqnarray}
m_{V,0}^2 &=& m_V^2+\delta m_V^2,\quad V=W, Z \nonumber \\
m_{f,0} &=& m_f+\delta m_f,
\end{eqnarray}
and the electric charge by
\begin{equation}
e_0=(1+\delta Z_e) e.
\label{eqn:re_e}
\end{equation}
Meanwhile, renormalization of the gauge fields are defined as
\begin{eqnarray}
W_0^\pm &=& (1+ \frac{1}{2}\delta Z_{WW}) W^\pm , \nonumber \\
\left(\begin{array}{c} Z \\ A \end{array}\right)_0
&=&
\left(
\begin{array}{cc}
  1+  \frac{1}{2}\delta Z_{ZZ}  &  \frac{1}{2}\delta Z_{ZA}  \\  \frac{1}{2}\delta Z_{AZ} &     1+  \frac{1}{2}\delta Z_{AA}
\end{array} 
\right)
\left(\begin{array}{c} Z \\ A \end{array}\right),
\label{eqn:re_za}
\end{eqnarray}
while for the fermion field, due to its left- and right-handed chirality, it has independent field renormalization constants:
\begin{eqnarray}
f^L_0 &=& (1+\frac{1}{2}\delta Z^L_f) f^L \nonumber \\
f^R_0 &=& (1+\frac{1}{2}\delta Z^R_f) f^R .
\end{eqnarray}
\item Higgs sector: the renormalization is performed in physics Higgs basis.
Renormalization of Higgs masses are defined as
\begin{equation}
m_{\phi,0}^2 = m_\phi^2+\delta m_\phi^2,\quad \phi=G^\pm, H^\pm, G^0, H^0, h^0 ,
\end{equation}
while renormalization of Higgs fields are given by
\beqa
\left(\begin{array}{c}H^0\\h^0\end{array}\right)_0
&=&
\left(\begin{array}{c c}
1 + \frac{1}{2}\delta Z_{H^0H^0} & \frac{1}{2}\delta Z_{H^0h^0}\\
\frac{1}{2}\delta Z_{h^0H^0} & 1 + \frac{1}{2}\delta Z_{h^0h^0}
\end{array}\right) 
\left(\begin{array}{c}H^0\\h^0 \end{array}\right),  
\nonumber \\
\left(\begin{array}{c}G^0\\A^0\end{array}\right)_0
&=&
\left(\begin{array}{c c}
1 + \frac{1}{2}\delta Z_{G^0G^0} & \frac{1}{2}\delta Z_{G^0A^0}\\
\frac{1}{2}\delta Z_{A^0G^0} & 1 + \frac{1}{2}\delta Z_{A^0A^0}
\end{array}\right) 
\left(\begin{array}{c}G^0\\A^0 \end{array}\right),  
\nonumber \\
\left(\begin{array}{c}G^\pm\\H^\pm\end{array}\right)_0
&=&
\left(\begin{array}{c c}
1 + \frac{1}{2}\delta Z_{G^\pm G^\pm} & \frac{1}{2}\delta Z_{G^\pm H^\pm}\\
\frac{1}{2}\delta Z_{H^\pm G^\pm} & 1 + \frac{1}{2}\delta Z_{H^\pm H^\pm}
\end{array}\right) 
\left(\begin{array}{c}G^\pm\\H^\pm \end{array}\right) .
\eeqa
For the mixing angles, they are defined as
\begin{eqnarray}
\alpha_0&=&\alpha + \delta \alpha , \nonumber \\
\beta_0 &=& \beta +\delta \beta .
\end{eqnarray}
We have skipped the renormalization of $\lambda_5$ as it is not needed in our calculation.  In order to expose the difference between the on-shell PTS and the other tadpole schemes, we also provide the renormalization of the tadpoles in the standard tadpole scheme, which are defined as given below: 
\begin{eqnarray}
T_{H^0,0}&=&T_{H^0} + \delta T_{H^0} , \nonumber \\
T_{h^0,0}&=&T_{h^0} + \delta T_{h^0} .
\label{eqn:ct_t1}
\end{eqnarray} 

In contrast, the on-shell PTS is based on the FJ tadpole scheme~\cite{Fleischer:1980ub}, where the shifts of the VEVs $v_{1,2}\rightarrow v_{1,2}+\Delta v_{1,2}$ are introduced. We can choose the values of $\Delta v_{1,2}$ order by order to make renormalized tadpoles vanishing in the perturbation expansion. In such a procedure, we are able to appropriately allocate the tadpole contributions and define all counter terms in a gauge-independent way.

It is noteworthy that in the on-shell PTS, there is no need to define tadpole counter terms given in Eq.~(\ref{eqn:ct_t1}) (i.e. $\delta T_{H^0,h^0}$ ) anymore. Instead, their roles are replaced by $\Delta T_{H^0, h^0}$ which are caused by the shift of the VEVs and are given in \cite{Krause:2016oke,Krause:2016xku} as 
\begin{eqnarray}
\Delta T_{H^0} &=&(c_\alpha \Delta v_1 + s_\alpha \Delta v_2 )  m_{H^0}^2 , \nonumber \\
\Delta T_{h^0}&=&( -s_\alpha \Delta v_1 + c_\alpha \Delta v_2 ) m_{h^0}^2 .
\label{eqn:ct_t2}
\end{eqnarray} 

To avoid confusion, we have deliberately changed the convention of $\delta T$ in \cite{Krause:2016oke,Krause:2016xku} to $\Delta T$ to indicate that the origin and meaning of $\Delta T$ is different from $\delta T$ given in Eq. (\ref{eqn:ct_t1}). Such a convention has also been adopted in \cite{Denner:2016etu}. 
\end{itemize}

\subsection{Renormalization conditions}  
$\Delta v_{1,2}$/$\Delta T_{H^0, h^0}$ will appear in self-energies and as well as vertices. It has already been revealed in Refs.~\cite{Krause:2016oke, Denner:2016etu} that the effect of $\Delta v_{1,2}$/$\Delta T_{H^0, h^0}$
can be easily included by adding possible tadpole diagrams to generic one-particle irreducible diagrams.  In this scheme, well-known on-shell conditions are used everywhere. The differences between well-known on-shell scheme and this scheme are : 
\begin{enumerate}
\item Tadpole counter terms $\delta T_{H^0,h^0}$ defined in Eq.~(\ref{eqn:ct_t1}) vanish in this scheme;
\item Self-energies $\Sigma(k^2)$ are replaced with $\Sigma^{\mathrm{tad}}(k^2)$ by adding possible tadpole contributions.
\end{enumerate}
It should be pointed out that the derivatives of self-energies $\Sigma^\prime(k^2)$ remain unchanged due to the fact that tadpole contributions are independent of external momentum $k$, thus most wave function renormalization constants are same. 

In the following, we present more necessary information about this on-shell PTS. For more details, please refer to Ref.~\cite{Krause:2016oke}.

In gauge and fermion sectors, same on-shell conditions as in the SM are used~\cite{Denner:1991kt}. Among all the renormalization constants in these sectors, three are different: $\delta m_W^2$, $\delta m_Z^2$ and $\delta m_f$. Only the first two are used in this work, they are given by:
\begin{equation}
\delta m_V^2 =\mathrm{Re} \Sigma^{\mathrm{tad},T}_{VV}(m_V^2), \quad V=W, Z 
\end{equation}

In Higgs sector, the renormalized self-energy of the Higgs field $\phi$ is the following finite combination of the unrenormalized self-energy
\begin{eqnarray*}
\widehat{\Sigma}_{\phi\phi}(k^2)&=&{\Sigma}_{\phi\phi}^{\mathrm{tad}}(k^2)-
\delta m^2_{\phi}+(k^2- m^2_{\phi})\, \delta Z_{\phi\phi} , \quad \phi=G^\pm, H^\pm, G^0, A^0, H^0, h^0\nonumber 
\end{eqnarray*}
while for the mixing one, it is
\begin{eqnarray*}
\widehat{\Sigma}_{\phi_1\phi_2}(k^2)&=&{\Sigma}_{\phi_1\phi_2}^{\mathrm{tad}}(k^2)
+\frac{1}{2}\delta Z_{\phi_1\phi_2}(k^2- m^2_{\phi_1}) +\frac{1}{2}\delta Z_{\phi_2\phi_1}(k^2- m^2_{\phi_2}),  \nonumber
\end{eqnarray*}
with $(\phi_1,\phi_2)=(G^\pm, H^\pm), (G^0, A^0)$ and $(H^0, h^0)$.

The renormalization conditions for the masses and wave functions in Higgs sector are given as:

\begin{itemize}
\item For the masses, on-shell conditions are used: 
\begin{equation}
\mathrm{Re}\widehat{\Sigma}_{\phi\phi}(m_\phi^2)=0, \quad \phi=H^\pm, A^0, H^0, h^0. 
\end{equation}
These conditions ensure that the physical masses of Higgs bosons are pole masses of corresponding renormalized propagators, from which we obtain
\begin{equation}
\delta m^2_{\phi}=\mathrm{Re}{\Sigma}_{\phi\phi}^{\mathrm{tad}}(m^2_{\phi}), \quad \phi=H^\pm, A^0, H^0, h^0
\end{equation}

\item For the wave functions, we require the residue of renormalized propagator of $\phi$ to be one:
\begin{equation}
\mathrm{Re}\widehat{\Sigma}^\prime_{\phi\phi}(k^2)|_{k^2=m_{\phi}^2}=0,  \quad \phi=G^\pm, H^\pm, G^0, A^0, H^0, h^0
\end{equation}
and the mixing self-energies vanish when one of the external particle is on shell:
\begin{equation}
\widehat{\Sigma}_{\phi_1 \phi_2}(m_{\phi_1}^2)=\widehat{\Sigma}_{\phi_1 \phi_2}(m_{\phi_2}^2)=0, \quad (\phi_1,\phi_2)=(G^\pm, H^\pm), (G^0, A^0), (H^0, h^0).
\end{equation}
From the equations we obtain
\begin{eqnarray}
\delta Z_{\phi\phi} &=&-\mathrm{Re}\Sigma^\prime_{\phi\phi} (m_{\phi\phi}^2), \quad \phi=G^\pm, H^\pm, G^0, A^0, H^0, h^0 \nonumber \\
\delta Z_{\phi_1 \phi_2} &=& \dfrac{2\mathrm{Re}\Sigma^{\mathrm{tad}}_{\phi_1\phi_2}(m_{\phi_2}^2)   }{m^2_{\phi_1}-m^2_{\phi_2}},  \quad (\phi_1,\phi_2)=(G^\pm, H^\pm), (G^0, A^0), (H^0, h^0) .
\end{eqnarray}
\end{itemize}

The renormalization of the mixing angles might be the most important part of this appendix. 
As the FJ tadpole scheme is applied, gauge-independent definitions for the counter terms $\delta\alpha$ and $\delta\beta$ are possible. In Ref.~\cite{Denner:2016etu}, 
$\overline{\mathrm{MS}}$ subtraction is used, while in this work, we follow the procedure in Ref.~\cite{Krause:2016oke} where the pinch technique~(see e.g. Ref.~\cite{Binosi:2009qm}) is applied:

\begin{itemize}
\item First, the pinched self-energies $\overline{\Sigma}$ are obtained with the help of the pinched technique. It is found out to a sum of two parts:
\begin{equation}
\overline{\Sigma}(k^2)=\Sigma^{\mathrm{tad}}|_{\xi=1} (k^2)+ \Sigma^{\mathrm{add}}(k^2),
\label{eqn:re_pt}
\end{equation}
where $\xi$ stands for the gauge fixing parameters $\xi_Z$, $\xi_W$ and $\xi_\gamma$ of the $R_\xi$ gauge.
As shown in Eq.~(\ref{eqn:re_pt}), the first part has the same form as the tadpole self-energies evaluated in the Feynman gauge, and the second one is an additional contribution which is explicitly independent of the gauge fixing parameter  $\xi$. 

\item The counter terms are then defined through above pinched self-energies with the on-shell scale, namely 
\begin{eqnarray}
\delta \alpha &=& \dfrac{\mathrm{Re}\overline{\Sigma}_{H^0h^0}(m_{H^0}^2)+\mathrm{Re}\overline{\Sigma}_{H^0h^0}(m_{h^0}^2)}{2(m_{H^0}^2-m_{h^0}^2)}, \nonumber \\
\delta \beta &=& -\dfrac{\mathrm{Re}\overline{\Sigma}_{G^0A^0}(0)+\mathrm{Re}\overline{\Sigma}_{G^0A^0}(m_{A^0}^2)}{2m_{A^0}^2} .
\end{eqnarray}
where the definition for $\delta\beta$ from CP-odd Higgs sector has been chosen.
\end{itemize}

More details about this on-shell PTS can be found in Ref.~\cite{Krause:2016oke}. For completeness, we list here the additional parts in the pinched self-energies as the end of this appendix:
\begin{eqnarray}
\Sigma^{\mathrm{add}}_{H^0h^0}(k^2)&=&
\dfrac{g^2s_{\beta-\alpha}c_{\beta-\alpha}}{32\pi^2c_W^2}
\biggl(k^2-\dfrac{m^2_{H^0}+m^2_{h^0}}{2}\biggr)
\biggl\{\biggl[B_0(k^2;m^2_Z,m^2_{A^0}) -B_0(k^2;m^2_Z,m^2_Z)\biggr]  
\nonumber \\ &&
+2c_W^2\biggl[B_0(k^2;m^2_W,m^2_{H^\pm})-B_0(k^2;m^2_W,m^2_W) \biggr]\biggr\} ,
\nonumber \\
\Sigma^{\mathrm{add}}_{G^0A^0}(k^2)&=&
\dfrac{g^2s_{\beta-\alpha}c_{\beta-\alpha}}{32\pi^2c_W^2}
\biggl(k^2-\dfrac{m^2_{A^0}}{2}\biggr)
\biggl[B_0(k^2;m^2_Z,m^2_{H^0}) -B_0(k^2;m^2_Z,m^2_{h^0})\biggr]  ,
\end{eqnarray}
where $B_0$ is the scalar two-point function~\cite{tHooft:1978jhc}.

\newpage
\bibliography{eezh_final}
\end{document}